  \documentclass[usenatbib,fleqn]{mn2e}
  
  \usepackage{amsmath}
  \usepackage{mathrsfs}
  \usepackage{amssymb}
  \usepackage{aas_macros}
  \usepackage{graphicx}
  \usepackage[dvipsnames]{xcolor}
  \usepackage{ulem}

  \usepackage[breaklinks,colorlinks,citecolor=blue]{hyperref}
  \usepackage[all]{hypcap} %Links go to figures. This breaks deluxetables; use +++\capstartfalse \capstarttrue around deluxeta

  \newif\iffigs
  \newif\iffigscl
  \newif\iffigstest
  \newif\iflabs
  
  \figstrue
  \figscltrue
  \figstestfalse
  
  \labsfalse

  \title[Estimating the gas expulsion time-scale]
  {How fast do young star clusters expel their natal gas?: Estimating the upper limit of the gas expulsion time-scale}
  \author[F. Dinnbier \& S. Walch]{Franti\v{s}ek Dinnbier$^{1,2}$\thanks{E-mail:dinnbier@ph1.uni-koeln.de} and 
  Stefanie Walch$^{1,3}$ \\
  $^{1}$I.Physikalisches Institut, Mathematisch-Naturwissenschaftliche Fakult\"{a}t,
  Universit\"{a}t zu K\"{o}ln, Z\"{u}lpicher Strasse 77, D-50937 K\"{o}ln, Germany \\
  $^{2}$Charles University in Prague, Faculty of Mathematics and Physics, Astronomical Institute, V Hole\v{s}ovi\v{c}k\'{a}ch 2, 180 00 Praha 8, Czech Republic \\
  $^{3}$Center for Data and Simulation Science, University of Cologne, Germany, www.cds.uni-koeln.de}

\begin{document}

\newcommand{\itii}[1]{{#1}}
\newcommand{\franta}[1]{\textbf{\color{green} #1}}
\newcommand{\itiitext}[1]{{#1}}

% newcommands in text, mainly abbreviations
\newcommand{\eq}[1]{eq. (\ref{#1})}
\newcommand{\eqp}[1]{(eq. \ref{#1})}
\newcommand{\eqb}[2]{eq. (\ref{#1}) and eq. (\ref{#2})}
\newcommand{\eqc}[3]{eq. (\ref{#1}), eq. (\ref{#2}) and eq. (\ref{#3})}
\newcommand{\refs}[1]{Sect. \ref{#1}}
\newcommand{\reff}[1]{Fig. \ref{#1}}
\newcommand{\reft}[1]{Table \ref{#1}}

\newcommand{\cii}[2]{(\cite{#1};\cite{#2})}
\newcommand{\ciii}[3]{(\cite{#1};\cite{#2};\cite{#3})}

% environments with another text format
\newcommand{\datum} [1] { \noindent \\#1: \\}
\newcommand{\pol}[1]{\vspace{2mm} \noindent \\ \textbf{#1} \\}
\newcommand{\code}[1]{\texttt{#1}}
\newcommand{\figpan}[1]{{\sc {#1}}}

% symbols in text
\newcommand{\sfe}{\mathrm{SFE}}
\newcommand{\nbdvi}{\textsc{nbody6} }
\newcommand{\nbdvid}{\textsc{nbody6}}
\newcommand{\flash}{\textsc{flash} }
\newcommand{\flashd}{\textsc{flash}}
\newcommand{\mum}{$\; \mu \mathrm{m} \;$}
\newcommand{\rop}{$\rho$ Oph }
\newcommand{\HT}{$\mathrm{H}_2$}
\newcommand{\Halpha}{$\mathrm{H}\alpha \;$}
\newcommand{\HI}{H {\sc i} }
\newcommand{\HII}{H {\sc ii} }
\renewcommand{\deg}{$^\circ$}

% math mode
\newcommand{\dd}{\mathrm{d}}
\newcommand{\acosh}{\mathrm{acosh}}
\newcommand{\sign}{\mathrm{sign}}
\newcommand{\cex}{\mathbf{e}_{x}}
\newcommand{\cey}{\mathbf{e}_{y}}
\newcommand{\cez}{\mathbf{e}_{z}}
\newcommand{\cer}{\mathbf{e}_{r}}
\newcommand{\ceR}{\mathbf{e}_{R}}

\newcommand{\llg}[1]{\log_{10}#1}
\newcommand{\pder}[2]{\frac{\partial #1}{\partial #2}}
\newcommand{\pderrow}[2]{\partial #1/\partial #2}
\newcommand{\nder}[2]{\frac{\dd #1}{\dd #2}}
\newcommand{\nderrow}[2]{{\dd #1}/{\dd #2}}

% Units
% all abbreviations for units begin with capital letter
% without a preceeding blank -- order is CGS, 'astrophysical' (pc/M_sun/Myr) and other
\newcommand{\Cmiii}{\, \mathrm{cm}^{-3}}
\newcommand{\Gcmii}{\, \mathrm{g} \, \, \mathrm{cm}^{-2}}
\newcommand{\Gcmiii}{\, \mathrm{g} \, \, \mathrm{cm}^{-3}}
\newcommand{\Kms}{\, \mathrm{km} \, \, \mathrm{s}^{-1}}
\newcommand{\Si}{\, \mathrm{s}^{-1}}
\newcommand{\Esi}{\, \mathrm{erg} \, \, \mathrm{s}^{-1}}
\newcommand{\Ee}{\, \mathrm{erg}}
\newcommand{\Yr}{\, \mathrm{yr}}
\newcommand{\Kyr}{\, \mathrm{kyr}}
\newcommand{\Myr}{\, \mathrm{Myr}}
\newcommand{\Gyr}{\, \mathrm{Gyr}}
\newcommand{\Msun}{\, \mathrm{M}_{\odot}}
\newcommand{\Rsun}{\, \mathrm{R}_{\odot}}
\newcommand{\Pc}{\, \mathrm{pc}}
\newcommand{\Kpc}{\, \mathrm{kpc}}
\newcommand{\Sd}{\Msun \, \Pc^{-2}}
\newcommand{\Ev}{\, \mathrm{eV}}
\newcommand{\Kk}{\, \mathrm{K}}
\newcommand{\Au}{\, \mathrm{AU}}
\newcommand{\Mas}{\, \mu \mathrm{as}}

  \date{Accepted 2020 August 18. Received 2020 August 18; in original form 2020 June 15}
  
  \pagerange{\pageref{firstpage}--\pageref{lastpage}} \pubyear{2020}
  
  \maketitle
  
  \label{firstpage}
  
  %%%%%%%%
  \begin{abstract}

Formation of massive stars within embedded star clusters starts a complex interplay
between their feedback, inflowing gas and stellar dynamics, which often includes close stellar encounters.
Hydrodynamical simulations usually resort to substantial simplifications to model embedded clusters.
Here, we address the simplification which approximates the whole star cluster by a single sink particle,
which completely neglects the internal stellar dynamics.
In order to model the internal stellar dynamics, we implement a Hermite predictor-corrector integration scheme
to the hydrodynamic code \textsc{flash}.
As we illustrate by a suite of tests, this integrator significantly outperforms the current leap-frog scheme,
and it is able to follow the dynamics of small compact stellar systems without the necessity to soften the gravitational potential.
We find that resolving individual massive stars instead of representing the whole cluster by a single energetic source has
a profound influence on the gas component: for clusters of mass less than $\lesssim 3 \times 10^3 \Msun$, it slows gas expulsion
by a factor of $\approx 5$ to $\approx 1 \Myr$, and it results in substantially more complex gas structures.
With increasing cluster mass (up to $\approx 3\times 10^3 \Msun$), the gas expulsion time-scale slightly decreases.
However, more massive clusters ($\gtrsim 5\times 10^3 \Msun$) are unable to clear their natal gas
with photoionising radiation and stellar winds only if they form with a star formation efficiency (SFE) of $1/3$.
This implies that the more massive clusters are either cleared with another feedback mechanism or
they form with a SFE higher than $1/3$.

  \end{abstract}
  %%%%%%%
  
  %%%%%%%%%
  %%%%%%%%

%%%%%%%%%%%%%%%%%%%%%%%%%%%%%%%%%%%%%%%%

\section{Introduction}

Modelling star formation from collapse of a molecular cloud inevitably leads to very high density gas and associated short time-scales that 
cannot be handled directly with current technology, but necessitates the approximation of sink particles for the simulation to continue \citep{Bate1995}. 
In more recent simulations, the sink particle method was extended to model stellar feedback, where the sink particle 
turns into a source of various forms of energy (e.g. ionising radiation, stellar winds, supernovae). 
For simulations including larger portions of gas (typically clouds more massive than $10^4 \Msun$), one sink 
particle usually represents the entire star cluster containing several hundreds or more stars 
(e.g. \citealt[][]{Dale2011,Walch2012,Dale2013,Hopkins2014,Geen2016,Rahner2017,Gatto2017,Hopkins2018,Kim2018}),
entirely neglecting the internal stellar dynamics of the star cluster.

However, star clusters manifest a multitude of dynamical processes, 
including mass segregation \citep[e.g.][]{Spitzer1969,Gunn1979,Bonnell1998,Hillenbrand1998,Baumgardt2003,McMillan2007,Subr2008,Allison2009,Moeckel2009,Parker2014,Spera2016,Dominguez2017,Pavlik2019}, 
close interactions between three to several bodies \citep[e.g.][]{Aarseth1971,Heggie1975,Tanikawa2012} 
resulting in hardening of binaries \citep{Heggie1975} and production 
of runaway stars \citep[e.g.][]{Fujii2011,Tetzlaff2011,Perets2012,Oh2015,Maiz_Apellaniz2018,Schoettler2019}. 
The dynamical cluster environment also impacts the stability and survivability of planetary systems \citep[e.g.][]{Spurzem2009,Shara2016,Cai2017}.

In addition, early feedback from young stars expels the gas which has not formed stars yet, terminating star 
formation and setting the star formation efficiency (SFE). 
In this work, we use the definition of SFE as $\sfe = M_{\rm cl}/(M_{\rm cl} + M_{\rm gas})$, where $M_{\rm cl}$ is 
the total stellar mass and $M_{\rm gas}$ the total gaseous mass within the same volume after the star forming event. 
As the gas is expelled, the gravitational potential of the star cluster shallows so that some stars can escape it or the whole cluster 
even disintegrates entirely depending mainly on the value of the SFE and the time-scale of gas expulsion 
\citep{Tutukov1978,Hills1980,Mathieu1983,Lada1984,Kroupa2001b,Geyer2001,Baumgardt2007}.
Even if no star gets expelled from the cluster by any of the aforementioned processes, 
massive stars occupy a non-zero volume around their birth-site.

These dynamical processes redistribute stars to substantially larger distances from their birth cluster than 
assumed in the approximation of the whole cluster by a single source. 
Massive stars located further away from their birth-sites impart their feedback preferentially to 
the loosely bound outer parts of the cloud, possibly affecting the cloud in a 
different way than if all massive stars were located in the single source at the cloud density centre. 

In order to capture the dynamics of stars in embedded star clusters, it is necessary to take into account 
the huge dynamical range of the orbital time-scales of the stars, which calls 
for a more efficient integrator than currently implemented in hydrodynamic codes. 
Until recent work of \citet{Wall2019}, who implement Hermite integrator in the \textsc{AMUSE} software 
framework \citep{Pelupessy2013}, it was mainly leap-frog and Runge-Kutta integrators which was used in both AMR and SPH 
schemes (e.g. \citealt{Bate2005,Federrath2010}). 

In order to undertake another step towards a more realistic simulations of star cluster formation, 
we implement a 4th order Hermite integrator for sink particles to the AMR code \flashd. 
We apply the integrating scheme to study the influence of resolving 
the dynamics of individual stars in embedded star clusters on the expulsion of the residual gas (i.e. the gas 
which has not been transformed to stars). 
We particularly aim on estimating the gas expulsion time-scale of this process. 

The rest of the paper is organised as follows. 
In \refs{sIntegrator}, we describe the new sink particle integrator module for \flashd,
which we develop to perform the intended simulations. 
Accuracy and performance tests of the module as well as the connection to other \flash modules 
is dealt with in \refs{sNumTests}. 
The gas expulsion from embedded star clusters is investigated in \refs{sMainSim}. 
We summarise our results in \refs{sSummary}.

%--------------------------------------------------------------------
\section{The integration scheme}

\label{sIntegrator}

The description of the integrator for gravitational force 
is divided to the description of the integration of sink particles (\refs{ssFSinks}), 
which is more sophisticated, and that of gas (\refs{ssFGas}), which is simple.

\subsection{Forces acting on sink particles}

\label{ssFSinks}

In embedded star clusters, a star (which is hereafter represented by a sink particle) 
is subjected to gravitational force
generated by the underlying gas distribution and also by other stars.
The gravitational field generated by gas is substantially smoother than the field generated by stars 
because of the smooth spatial extent of the former and the compactness and large velocities of the latter 
\footnote{
This is a general property, which follows from typical density of gas and stars and the Poisson equation, and 
it is independent on the spatial distribution of gas or stars within the system.
}.
Stars, and particularly massive stars, are often packed in compact volumes near centres of young star 
clusters (either as the result of in situ formation or due to dynamical mass segregation), 
where they strongly interact forming dynamically unstable systems composed 
of several bodies \citep{PflammAltenburg2006,Allison2011,Tanikawa2012}. 
These systems rapidly evolve forming tight binaries often with binary recoils, while other 
stars are ejected \citep{Heggie1975,PflammAltenburg2006,Oh2015}. 
Moreover, massive stars are exclusively formed in binaries, with many of them being of short orbital periods 
(\citealt{Sana2012,Moe2017}; around 50\% of O stars have orbital periods shorter than 100 days). 
This implies that the dynamical time-scale for the integrating scheme must be able to capture a huge dynamical 
range, where the shortest time-steps are of the order of a fraction of an hour. 
In contrast, the gravitational field generated by the gas changes on a substantially longer time-scale, 
which corresponds to accretion inflows or dispersal due to feedback.

The two different time-scales for the interaction between star-star and star-gas motivate 
us to adopt the spirit of the Ahmad-Cohen method \citep{Ahmad1973}, 
where stars are split to two groups based on their physical proximity and each group is integrated by 
its own time-step; the irregular force with shorter time-step originates from the closer group of stars, while the 
regular force with larger time-step originates from the rest of the cluster. 
However here, we split the gravitational force according to the kind of matter which generates the gravitational field; 
the irregular force originates from stars, while the regular force originates from gas.   
In our implementation, the split of the gravitational force does not take into account the physical proximity. 

An attractive feature of splitting the force in this way is the speed of code execution because calculating the force 
star-gas requires communication between all processors, and is therefore time consuming. 
In contrast, calculating the force star-star can be done locally as all the information about stars consumes a rather small 
amount of memory. 
Moreover, the predictor nature of the Hermite integrator enable us to extrapolate the smoothly varying force star-gas while 
the rapidly changing force star-star can be evaluated many times with short time-steps. 
This presents significant advantage of the predictor scheme over the commonly used leap-frog scheme.
We set the duration of the regular time-step $\Delta t_{R}$ to be the hydrodynamical time-step, 
so the regular force is evaluated only once per the hydrodynamical time-step. 
All sink particles have the same $\Delta t_{R}$.

\subsubsection{Irregular time-steps: forces due to sink particles}

The irregular time-steps $\Delta t_{I}$ are calculated individually for each particle 
according to the standard Aarseth formula \citep{Aarseth1985,Aarseth2003},
\begin{equation}
\Delta t_{I} = \sqrt{\frac{\eta (|\mathbf{F}_{I}| |\mathbf{F}_{I}^{(2)}| + |\mathbf{F}_{I}^{(1)}|^2)}
{(|\mathbf{F}_{I}^{(1)}| |\mathbf{F}_{I}^{(3)}| + |\mathbf{F}_{I}^{(2)}|^2)}},
\label{edtAarseth}
\end{equation}
where $\mathbf{F}_{I}$ is the irregular force acting on given star, and the numbers in round brackets 
indicate the order of the time derivative. 

The integration of sink particles moving in gaseous potential is realised as follows. 
First, the irregular time-steps are quantised by factor of 2 as in the usual block time-step 
method \citep{Aarseth2003}, so 
the current hydrodynamic half-time-step is of length $1$ in the quantised units, and the $j$-th 
time-step is of length $2^{-(j-1)}$. 
Index $j$ runs from 1 to 40, which corresponds to a dynamical range in $\Delta t_{I}$ of the order of $1$ to $10^{12}$. 
During the predictor part of the integrator, 
the positions and velocities are extrapolated by using the force and its time derivatives. 
The force is the sum of the irregular force $\mathbf{F}_{I}$, which is evaluated directly, and 
the regular force $\mathbf{F}_{R}$ (which was obtained at the end of the previous time-step) extrapolated to current time. 
We experiment with predictor to order $\mathbf{F}^{(1)}$, and also to order $\mathbf{F}^{(3)}$. 
When the predicted distances are determined, the new force derivatives are calculated by direct 
summation (i.e. not by an octal tree) over the other $N - 1$ sink particles, whereupon the new positions are corrected. 
Then, the new irregular time-step is calculated from formula (\ref{edtAarseth}) and quantised. 
This procedure continues until all sink particles are integrated to current time, 
i.e. advanced by time-step $1$ in quantised units.
The advantage of this calculation is that it can be done locally on each processor without any need of communication. 

\subsubsection{Regular time-steps: forces due to gas}

After the particles are advanced by irregular time-steps to the current simulation time, 
the gravitational force due to gas on sink particles is calculated by the standard octal tree method 
\citep{Barnes1986,Salmon1994}. %, where we use the particular implementation to \flash described by \citet{Wunsch2018}. 
The information about the distribution of gas within the whole computational domain must be 
obtained via communication with the other processors, so it is desirable that this time consuming operation 
is done only once per half time-step. 
During the communication, the information about gas distribution is also used for calculating the 
force acting on gas, which is described in \refs{ssFGas}.
In addition to standard octal tree method, we need to calculate the force to order $\mathbf{F}_{R}^{(1)}$. 
To evaluate the term $\mathbf{F}_{R}^{(1)}$, we consider new block node property: 
mass-weighted velocity $v_{node}$. 
This property is propagated, communicated and evaluated in the usual way as described in \citet{Wunsch2018} according to the geometric 
multipole acceptance criterion with opening angle $\theta$. 
According to this criterion, blocks are recursively split 
in eight child blocks (each of the same volume) until they are seen from the sink particle under an angle smaller then $\theta$. 
The angle $\theta$ is a constant during the whole simulation.

Then, the force derivative is calculated as
\begin{equation}
\mathbf{F}_{R}^{(1)} = G m_{s} \sum_{k = 1}^{N_{nodes}} m_{node,k} \left( -\frac{\mathbf{v}_{k}}{r_{k}^3} + 
\frac{3 (\mathbf{r}_{k} \cdot \mathbf{v}_{k}) \mathbf{r}_{k}}{r_{k}^5} \right) ,
\label{efdot}
\end{equation}
where the index $k$ runs over all accepted nodes. 
The relative distance and velocity between the sink particle and node centre 
is denoted $\mathbf{r}_{k} = \mathbf{r}_{s} - \mathbf{r}_{node,k}$ 
and $\mathbf{v}_{k} = \mathbf{v}_{s}-\mathbf{v}_{node,k}$, respectively. 
The mass of a node and sink particle is denoted $m_{node,k}$ and $m_{s}$, respectively.

In addition to configurations with isolated boundary conditions for gravity, the code is able to calculate 
also configurations with periodic and mixed boundary conditions for gravity using the Ewald method or 
its modifications as described in \citet{Wunsch2018}. 
We implement this property also for sink particles. 
This allows to study systems whose computational domain is surrounded by an infinite number of its periodic copies in one, two or three spatial directions. 
In these cases, sink particles experience the gravitational force also from all the other periodic copies of the computational domain. 
For example, these configurations might be suitable for modelling filaments \citep{Clarke2016} or stratified boxes within galactic discs \citep{Walch2015a}.
This means that if periodic or mixed boundary conditions are adopted, 
sink particles "feel" the gravitational force also from all the other periodic copies of the computational domain.

After the derivatives of the regular force are obtained for each sink particle, the correction 
for the regular force is calculated, and the sink particle position at the current time evaluated. 

\subsubsection{The softening radii}

In order to take into account the strong dynamical encounters between stars, 
the softening radius $r_{\rm soft,ss}$ for sink-sink interaction 
and sink-gas interaction $r_{\rm soft,sg}$ can be set independently, 
with the former one being typically smaller. 
We test (\refs{sssSmallSyst}) that the numerical scheme is able to cope with softening 
radii $r_{\rm soft,ss}$ as small as the stellar radius. 
We use the recommended value of the sink-gas softening radius 
$r_{\rm soft,sg}$ to be $2.5$ grid cell size at the highest refinement level 
as recommended by \citet{Federrath2010}; too small value of $r_{\rm soft,sg}$ is dangerous because it could 
cause a strong interaction between an extended gas element and a sink particle, which is artificial. 

The code also enables two groups of sink particles with two different softening radii, $r_{\rm soft,ss,1}$ and $r_{\rm soft,ss,2}$, 
which is intended for projects where a handful of massive stars 
(the first group), which dominates stellar dynamics 
and feedback, is integrated accurately at a high CPU cost, while much larger second group representing 
low mass stars does not evolve dynamically so fast and can be integrated with a larger softening radius. 
Tests containing sink particles of two different softening radii are presented in \refs{sssTwoSoftRadsEC} 
and \ref{sssTwoSoftRadsGE}. 

We adopted the particular form of the softening potential from eq. 21 of \citet{Monaghan1985} (see also 
appendix A of \citealt{Federrath2010}). 
Unlike many other softening potentials (e.g. the Plummer softening), 
this potential continuously transforms to the $1/r$ potential at $r_{\rm soft,ss}$, so the stellar dynamics is 
exact outside the softening radius ($r > r_{\rm soft,ss}$). 
For a reader interested in implementing the softening potential in a Hermite scheme, we list the 
corresponding formulae in Appendix \ref{sAppendixA}.

\subsection{Forces acting on gas}

\label{ssFGas}

%The gaseous component of the embedded cluster is subjected to the gravitational field 
%from stars and gas. 
Unlike stellar dynamics which is dominated by the gravitational force, 
gaseous dynamics is influenced by other forces, for example thermal and ram pressure gradients, and radiation pressure 
from the newly formed stars. 
Since these non-gravitational forces are typically evaluated with large uncertainties, 
the criteria for calculating the gravitational force for gas are also less strict than those for sink particles. 
Accordingly, the gravitational acceleration due to gas is calculated by the standard octal tree method \citep{Wunsch2018}. 

The gravitational acceleration due to sinks is calculated as follows: 
First, the mass of sink particles is distributed on grid according to formula (\ref{eKernDens}). 
Then, the mass of sink particles is propagated via the octal tree as a new node property, and during force 
evaluation added to the gaseous mass within the particular block. 

This time consuming calculation is done once per one hydrodynamical half-time-step. 
At that instant, the tree of the gas distribution is also utilised for calculating 
the force due to gas acting on sink particles as described in the previous section.

\section{Accuracy and performance tests}

\label{sNumTests}

We take the usual steps and require that before addressing more complex problems, 
the code must succeed in less complex situations. 
First, in \refs{ssOnlySinks}, we study the accuracy of the implementation of the Hermite integrator with softening 
employed on simple problems including from two to ten thousand sink particles with negligible 
influence of the gaseous potential. 
In these settings, \flash behaves like a simple N-body integrator without regularising techniques. 
Then, in \refs{ssSinksOnGas}, we study the accuracy of gravitational interaction between sink 
particles and gaseous spheres including gravitational collapse and accretion. 
When comparing to \nbdvid, we refer to the new \flash integrator as \textsc{HermiteSink}, 
which is the name of the new module.

\subsection{Sink particles interacting with themselves}

\label{ssOnlySinks}

To suppress the gravitational influence of the gas, we set its density to $\rho = 10^{-40} \Gcmiii$ 
for all tests in this section.
This enables us to easily verify the simulation by checking the evolution of the energy error because 
negligible energy is exchanged with the gaseous component. 
Throughout this section, the relative energy error $\Delta e(t)$ at time $t$ is defined as
\begin{equation}
\Delta e(t) = |E(t) - E(t=0)|/|E(t=0)|,
\label{eRadRatio}
\end{equation}
where $E(t)$ is the total mechanical energy (i.e. the kinetic and potential energy) 
of all the sink particles in the system. 

\subsubsection{Two body problem}

\label{sssTwoBody}

We use a two body problem to show that the implementation behaves like a 4th order integrator, 
i.e. that the energy error decreases with the size of a time-step $\dd t$ as $(\dd t)^{4}$. 
The bodies are of non-equal masses, with the primary and secondary mass $1 \Msun$ and $3 \times 10^{-6} \Msun$, 
respectively. 
Their orbit is eccentric with $\epsilon = 0.9$ and semi-major axis $a = 1 \Au$. 

The time dependence of the energy error for time-steps calculated with $\eta = 0.01$, 
$\eta = 0.001$ and $\eta = 0.0001$ (cf. \eq{edtAarseth}) is shown in the left panel of Figure \ref{fTwoBody}. 
The increase of $\eta$ by factor of 10 means a time-step larger by a factor of $\sqrt{10}$, 
and an energy error larger by a factor of 100. 
This scaling is indicated by the thin black lines, which represent the 
error for the model with $\eta = 0.0001$ multiplied by 100 and 10000, respectively. 
The scaled errors are of the same order as the error of models with $\eta = 0.001$ and $\eta = 0.01$.
Note that the model with $\eta = 0.01$ has 342 time-steps per orbit. 
Qualitatively, the error for a two body problem of the same eccentricity is listed in \citet[][his table 13.1]{Aarseth2003}, 
which is $\Delta e \approx 7.8 \times 10^{-7}$. 
Our implementation has an error of the same order of magnitude, 
$\Delta e \approx 16 \times 10^{-7}$ when averaged over 1000 orbits. 
We attribute the slightly worse error in our implementation to the large difference in the masses of the bodies. 

The previous test is calculated with predictor 
to the order $\mathbf{F}^{(1)}$ ($\mathbf{F} = \mathbf{F}_{I} + \mathbf{F}_{R}$). 
The right panel of Fig. \ref{fTwoBody} shows the energy error for the same test but 
with predictor to the order $\mathbf{F}^{(3)}$ (only the star to be advanced is predicted to the high order, 
the other stars are predicted to the order $\mathbf{F}^{(1)}$). 
The higher order predictor results in more accurate calculations (the  
orbits with $\eta = 0.01$ and $\eta = 0.001$ have by factor of $10$ smaller $\Delta e$ at the end of the simulations). 
Since the costs of using a higher order predictor are small, we use this predictor as a default option of the integrator. 

\iffigscl
\begin{figure*}
\includegraphics[width=\textwidth]{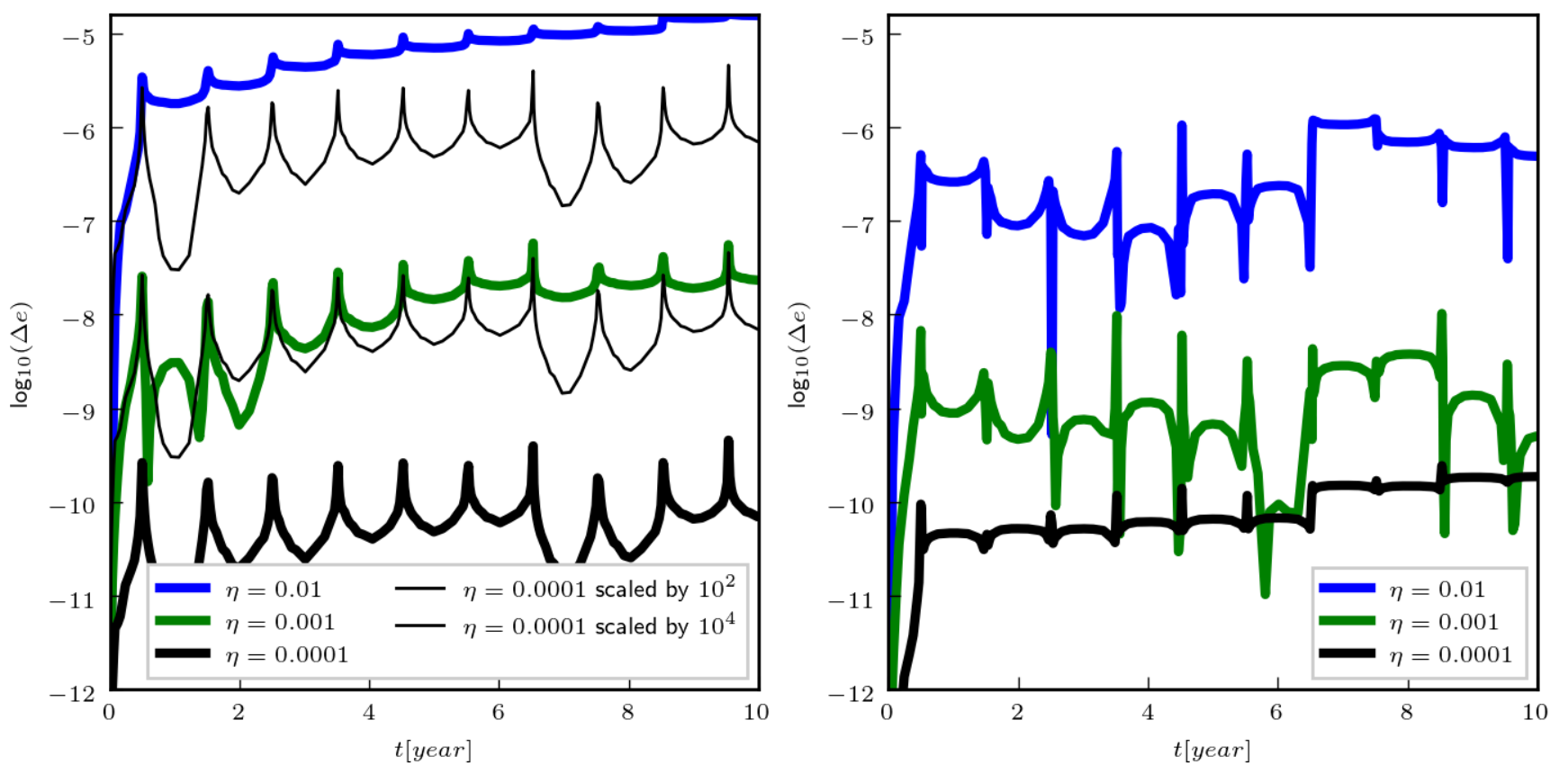}
\caption{Evolution of the relative energy error $\Delta e$ as a function of time for 
a two body problem of unequal masses ($M_{\rm 2}/M_{\rm 1} = 3\times 10^{-6}$) 
with eccentricity $\epsilon = 0.9$ and the semi-major axis of $1 \Au$. 
The orbits are integrated by the Hermite predictor-corrector scheme 
with three different values of the constant $\eta$ (c.f. \eq{edtAarseth}) 
as indicated by the thick lines.
\figpan{Left panel:} Predictor to the order $\mathbf{F}^{(1)}$. 
The thin black lines show the error for the model with $\eta = 0.0001$ scaled so that 
it corresponds to the expected errors of a 4th order integrator with $\eta = 0.001$ and $\eta = 0.01$.
\figpan{Right panel:} Predictor to the order $\mathbf{F}^{(3)}$.}
\label{fTwoBody}
\end{figure*} \else \fi

%\franta{This paragraph used to be a subsection with a plot, but I think it is unnecessary.}
We use the same initial conditions to test the implementation of the softening potential. 
For this test, we set $r_{\rm soft,ss} = (1+\epsilon) a/1.2$, which ensures that the bodies 
approach each other at a distance $r < r_{\rm soft,ss}/2$ at the pericentre, and 
recede from each other at a distance $r > r_{\rm soft,ss}$ at the apocentre. 
Such a trajectory covers all the three different functional forms of the softened potential \eqp{eKernPot}. 
After $10$ orbits in the softened potential, the relative energy error is $\log_{10} (\Delta e) = -6.04$ ($\eta = 0.01$), 
which is of the same order as $\Delta e$ for the potential without softening (right panel of \reff{fTwoBody}) 
after the same number of orbits, indicating that the softening potential is implemented correctly in the code.

%\subsubsection{The softening potential}
%
%The trajectory of the particles is chosen so that it starts outside the softening radius, 
%and crosses both $r_{\rm soft}$ and $r_{\rm soft}/2$, where the prescription of the softening potential changes 
%(c.f. \eq{esoftPot}) to test that all parts of the function describing the softening were implemented 
%correctly. 
%The orbital parameters of the particles and their masses are the same as for the two body test, \refs{sssTwoBody},
%the softening radius is slightly smaller than the apocentric distance, 
%$r_{\rm soft} = (1+\epsilon) a/1.2$. 
%The first ten years of evolution are shown in the right panel of Fig. \ref{fSoft}. 
%The energy error during 1000 orbits is shown in the left panel of Fig. \ref{fSoft}, 
%which corroborates energy conservation during many passages through the softening kernel.

%\iffigscl
%\begin{figure*}
%\includegraphics[width=\textwidth]{soft}
%\caption{\figpan{Left panel:} Evolution of the relative energy error $\Delta e$ for two 
%particles orbiting in their softened potentials. 
%\figpan{Right panel:} Trajectory of the lighter particle (green) in the potential of 
%the more massive particle (blue) during the first ten years of orbit. 
%The softening radius of the more massive particle is indicated by the red circle.} 
%\label{fsoft}
%\end{figure*} \else \fi
%

\subsubsection{Mixed boundary conditions}

\label{sssMixed}

If the astrophysical system in question possesses a symmetry, it is often beneficial to 
use this fact to facilitate the calculations. 
For example, a box of side lengths $L_x$, $L_y$ and $L_z$ placed on the Galactic disc
can be, to some approximation, translated in the $x$ direction by distance $L_x$ or 
in the $y$ direction by distance $L_y$ (direction $z$ is normal to the galactic plane) 
without changing the gravitational force inside the box. 
In this approximation, the gravitational force of the whole galactic 
disc is calculated as if the computational domain were copied in directions $x$ and $y$ 
to infinite distances, and the gravitational force of each of the copy summed together. 
Thus, this galactic setup has mixed boundary conditions (i.e. periodic in directions $x$ 
and $y$, and isolated in the direction $z$). 
Similarly, a piece of a straight filament (with symmetry axis in direction $x$) is 
approximately symmetric in respect to translation by the distance $L_x$. 
Thus, the filamentary setup has another type of mixed boundary conditions, 
periodic in the direction $x$ and isolated in the other directions $y$ and $z$.
Some systems can be also approximated with periodic boundary conditions in all three directions. 

For computational domains with periodic, or mixed 
boundary conditions, it is advantageous to speed up the convergence of the 
gravitational force calculation by the Ewald method \citep{Ewald1921,Springel2005,Wunsch2018}. 
We connected the sink particles to the Ewald method, so that not only gas, but also sink 
particles feel the gravitational force from the periodic copies of the computational domain. 
Likewise, not only gas, but also sink particles are replicated for the purpose of 
calculating the gravitational force.

To test the implementation of mixed boundary conditions (BCs), we set the configuration as follows.
The box is a cube of a side-length $L_x = L_y = L_z = 1 \Pc$ with periodic boundary conditions in directions $x$ 
and $y$, and with isolated BCs in the direction $z$. 
There is a massive sink particle of mass $1 \Msun$ at $(0,0,L_x/2)$, and a test particle 
of negligible mass at $(L_x/2,L_x/2,L_x/2 + L_x/10)$. 
The displacement of the test particle by $L_x/10$ makes the particle to oscillate in 
the $z$ direction through the plane of symmetry, which is at $z = L_x/2$. 
In the approximations of small displacements, 
the $z$ motion of the test particle follows the equation of harmonic oscillator, 
i.e. $z(t) = L_x/2 + (L_x/10) \cos(2\pi t/T)$ with period $T = 22.9 \Myr$.

Figure \ref{ftestMixed} shows the evolution of the $z$ displacement of the test particle (black line), 
and the harmonic approximation (red line). 
The measured period of oscillation, $23.2 \Myr$, 
is very close to the analytic value of $22.9 \Myr$. 
Sink particles are connected to the Ewald method also for computational domains with periodic boundary 
conditions in one direction and with isolated boundary conditions in the other two directions, and for 
computational domains with boundary conditions periodic in all the three directions. 

\iffigscl
\begin{figure}
\includegraphics[width=\columnwidth]{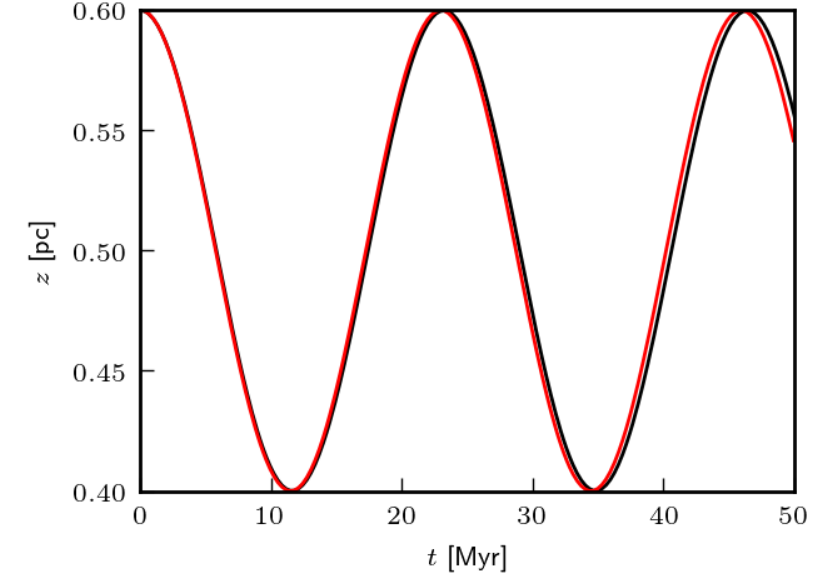}
\caption{The vertical position $z$ of the test particle 
oscillating through the plane of symmetry at $z = 0.5 \Pc$ (black line). %; the test is described in \refs{sssMixed}). 
The analytical solution is shown by the red line.}
\label{ftestMixed}
\end{figure} \else \fi

\subsubsection{Star clusters and their performance test}

\iffigscl
\begin{figure*}
\includegraphics[width=\textwidth]{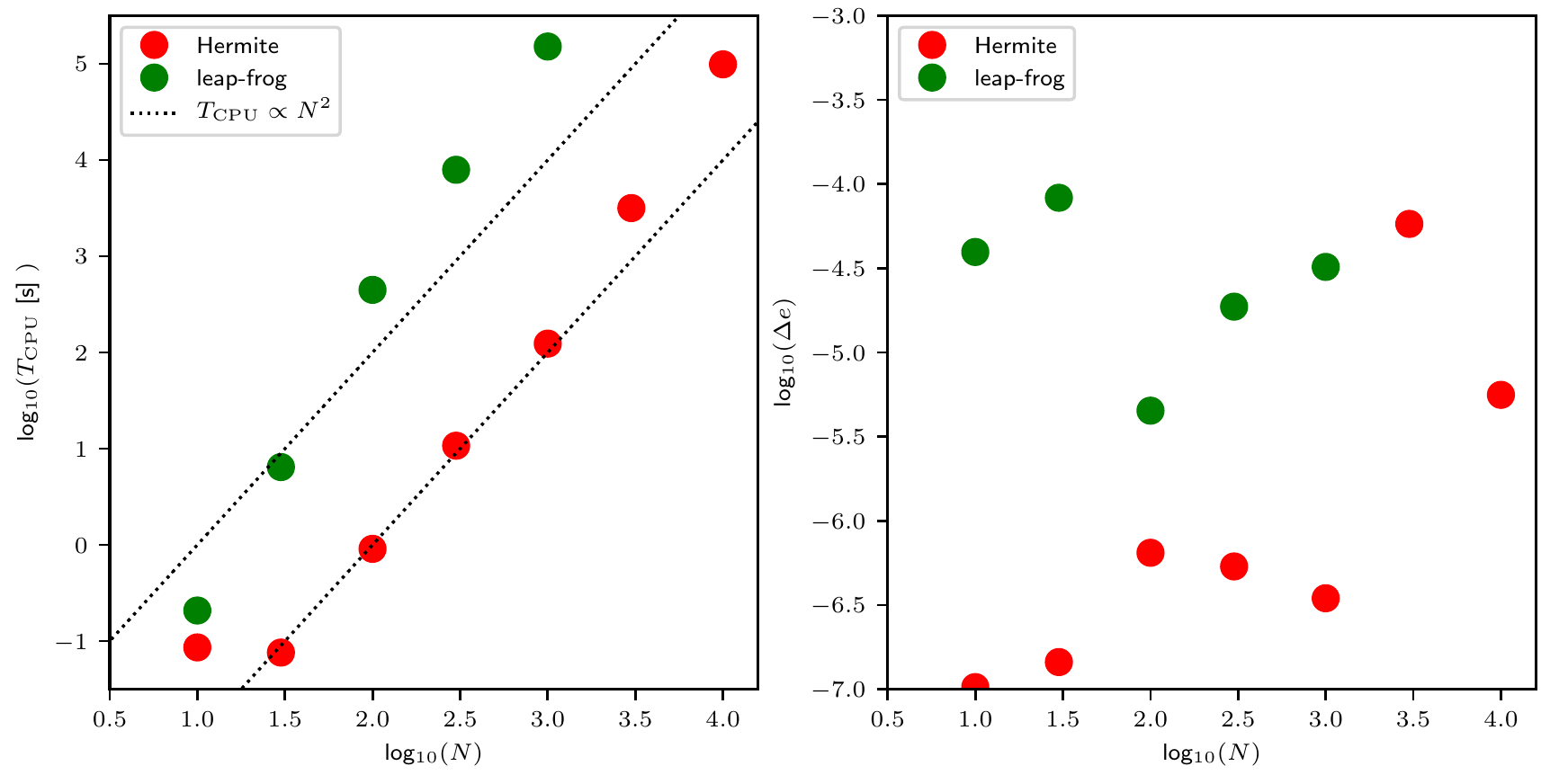}
\caption{\figpan{Left panel:} Scaling of the walk-clock time $T_{\rm CPU}$ 
as a function of the number of particles $N$ within a cluster over 10 half-mass radius crossing times. 
Results of the leap-frog scheme and the Hermite scheme are plotted by green and red circles, respectively.
The theoretical scaling $T_{\rm CPU} \propto N^2$ is indicated by dotted lines. 
\figpan{Right panel:} The dependence of the relative energy error $\Delta e$ 
on the number of particles $N$ for the two integrators.
The Hermite scheme is not only $\sim 1000 \times$ faster than the leap-frog, 
but it is also more accurate.}
\label{fClusters}
\end{figure*} \else \fi

We investigate the CPU costs and energy error in simulations of 
star clusters containing from $N = 10$ to $N = 10^4$ stars, 
where each star is represented by one sink particle. 
At the beginning, the density distribution of the clusters corresponds to 
the Plummer model with Plummer parameter $a_{\rm Pl} = 0.1 \Pc$. 
The stellar velocities are isotropic and chosen so that the cluster is in virial equilibrium \citep{Aarseth1974b}, 
The softening radius is $1 \Au \approx 5 \times 10^{-6} \Pc$. 
The masses of the stars are randomly drawn from the Salpeter IMF \citep{Salpeter1955} within 
the mass range $(0.1 \Msun, 120 \Msun)$
\footnote{This overproduces low mass stars in contrast to modern studies of the IMF \citep{Kroupa2001a,Chabrier2003}, 
but the system described in this section aims 
at testing the implementation rather than realistic models of star clusters.}
.
The clusters are integrated for 10 half-mass radius crossing times, 
which corresponds to $12 \Myr$ for the 10 particle cluster, and to $0.4 \Myr$ for the $10^4$ particle cluster. 

The scaling of the code with increasing number of particles is shown in the left panel of Fig. \ref{fClusters}. 
The same clusters take approximately $1000 \times$ more time to be calculated with the leap-frog scheme (green circles) 
than with the Hermite scheme (red circles). 
Note that the leap-frog uses the same time-step for all particles, which likely makes up the most of 
the difference. 
The dashed lines indicate the expected scaling $T_{cpu} \propto N^2$ for a softened problem of $N$ bodies. 
While the Hermite scheme follows quite closely the expected scaling for clusters with $N < 3 \times 10^3$, 
the computational demands for the leap-frog tend to increase steeper. 
The deviation from the ideal scaling for the clusters with $N = 10$ is probably due to the inability of the CPU time 
measuring routines to measure accurately the relevant calls, which are very short in this case.
%The number of time-steps per particle during the whole calculation is shown in \reff{fstepsPerPart}.

The relative energy error was calculated at the end of each simulation, and the results 
are plotted in the right panel of Fig. \ref{fClusters}. 
The plot shows that although the leap-frog is $\sim 10^3 \times$ slower, it 
is even less accurate than the Hermite scheme.

\subsubsection{Tightly bound systems free of softening}

\label{sssSmallSyst}

An important feature of massive stars is that they often form temporal systems consisting of several bodies 
(an example is the Trapezium at the centre of the Orion star forming region), 
which are dynamically unstable, with binary recoils, binary hardening \citep{Heggie1975}, 
and production of runaway stars \citep{PflammAltenburg2006,Fujii2011,Tanikawa2012,Wang2019}. 
It is desirable that these features are correctly implemented also in a hydrodynamical code so that 
stellar dynamics and stellar feedback are included self-consistently. 

In order to test the integrator on these systems, we perform a statistical analysis of $100$ simulations 
of the same compact system consisting of $10$ massive stars. 
The system is a Plummer sphere with $a_{Pl} = 0.025 \Pc$ in virial equilibrium at the beginning, 
and the stars are generated from the Kroupa IMF 
in the mass range ($15 \Msun$, $40 \Msun$). 
To treat stellar dynamics as accurately as possible, we set the softening 
radius to be two times the radius of a B0 star, which we take 
to be $7.5 R_{\odot}$ (c.f. table 3.13 in \citealt{Binney1998}), so
$r_{\rm soft,ss} = 15 R_{\odot} = 3.0 \times 10^{-7} \Pc$. 
This means that stellar dynamics is reproduced correctly until two massive stars touch each other; 
In other words, the softening is decreased to such an extent that it plays role only during direct stellar collisions.
The initial conditions differ only in the random number used for generating stellar positions, velocities and masses. 
To benchmark our implementation, we integrated exactly the same clusters by the state-of-the art code \nbdvid.

\iffigscl
\begin{figure*}
\includegraphics[width=\textwidth]{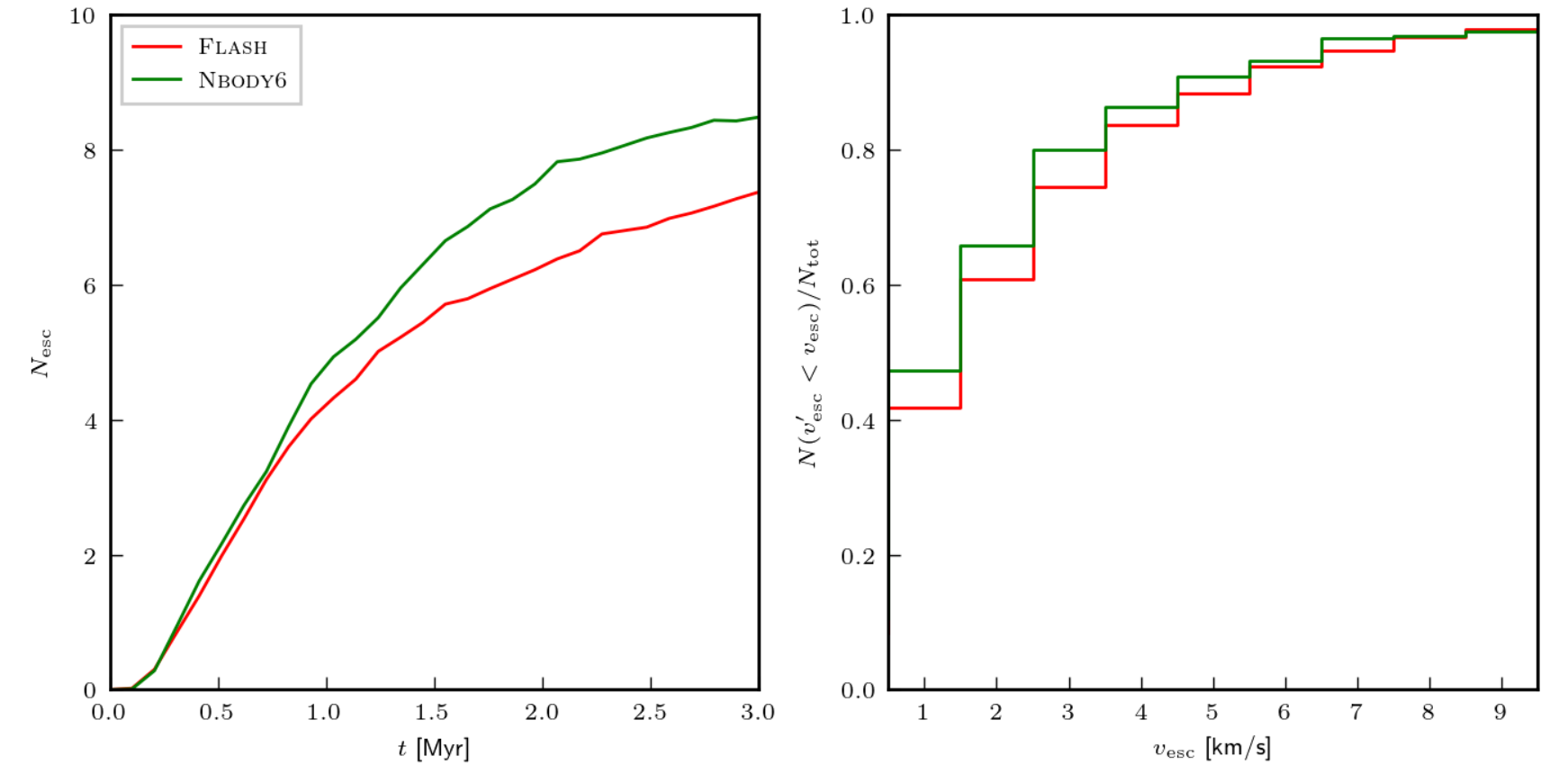}
\caption{
Comparison of the \textsc{HermiteSink} \textsc{flash} module (red lines) with the sophisticated 
code \textsc{Nbody6} (green lines) on the test problem of \refs{sssSmallSyst} (small compact star clusters). 
The total number of stars in the system is $N_{\rm tot} = 10$. 
The results are averages over $100$ different realisation of the same model.
\figpan{Left panel:} The number $N_{\rm esc}$ of stars which have escaped from the 
compact stellar system by time $t$. 
\figpan{Right panel:} Cumulative velocity distribution of escaping stars.}
\label{fScatter}
\end{figure*} \else \fi

%For the analysis, we take into account only the Nbody6 models which terminated successfully, and the Flash models 
%which were calculated with relative energy error smaller than $10^{-5}$. 
Left panel of \reff{fScatter} compares the number of stars $N_{esc}$
which escaped from the system as  a function of time. 
An escaper is defined as a star which is projected further away than $0.5 \Pc$ from the cluster density centre. 
Both codes are almost identical to each other by $1 \Myr$, with \nbdvi producing 
slightly more escapers after that (8.5 vs 7.4 at $t = 3 \Myr$). 

Another quantity of interest is the velocity distribution of escaping stars. 
We consider as escapers all stars which are gravitationally unbound at time  $t = 3 \Myr$; 
we calculate the binding energy for every star at that time, and if the energy is positive, we take 
the star as an escaper. 
Then, its escape velocity $v_{\rm esc}$ is calculated relative to the velocity of the mass centre of the system. 
The \textsc{HermiteSink} module and \nbdvi produce 4.2 and 4.8 escapers 
per simulation, respectively. 
The velocity distribution of escaping stars calculated by the \textsc{HermiteSink} module, 
as shown in the right  panel of \reff{fScatter} (red line), 
is very close to the velocity distribution of escaping stars as calculated by \nbdvid (green line).
Also, the number of stars with escaping velocity in excess of $10 \Kms$ is close between the codes 
(0.09 and 0.12 per model for the \textsc{HermiteSink} module and \nbdvi, respectively), 
demonstrating that the \textsc{HermiteSink} module is capable of producing correct number of fast ejectors. 

Thus, the \textsc{HermiteSink} is capable of treating strong stellar encounters very well when used 
with softening radius of the order of stellar radius.
Obviously, the cost of this brute force approach is the CPU time, which in these models is $60 \; \rmn{s}$ 
per simulation, while it is $0.8 \; \rmn{s}$ for \nbdvid. 
The median relative energy error from these simulations is  $7\times 10^{-7}$ for the 
\textsc{HermiteSink} module and $4\times 10^{-3}$ for \nbdvid.
With increasing number of stars, the CPU demands increase substantially, but this test demonstrates that it is possible 
to accurately model a dynamics of small stellar systems for several Myrs, 
which is long enough to simulate their embedded phase.

\subsubsection{Clusters with two different softening radii for sink particles: Energy conservation}

\label{sssTwoSoftRadsEC}

%Two different models: clusters without gas - energy error
%Clusters with gaseus potential - half-mass radii

Since in star cluster formation models the main focus is on the dynamics of the most massive stars, 
which dominate feedback, 
the dynamics of lower mass stars ($m \lesssim 9 \Msun$) leaves a substantial room for approximations. 
It is desirable to decrease the number of lower mass stars 
as they significantly outnumber their higher mass counterparts (aprox. 1:300 
for a randomly sampled Kroupa IMF in the interval $(0.01 \Msun, 120 \Msun)$). 
Lower mass stars also dominate the mass of the cluster, comprising 
$80.5$ \% of total cluster stellar mass for the adopted IMF. 
The only influence of lower mass stars on the gas and massive stars is due to their gravitational force, 
so the approximation of the lower mass stars should model their gravitational force correctly. 

We approximate many lower mass stars with one sink particle. 
However, this implies that the reduced number of particles has shorter relaxation time-scale than a real cluster. 
Particularly, clusters with massive stars usually contain more than several hundreds of stars in total, making 
the relaxation time of the low mass stars to be of the order of $\approx 3 \Myr$, and longer for 
more massive clusters. 
This means that the relaxation time-scale is comparable, or longer than the life-time of embedded 
star clusters. 
Accordingly, we prevent relaxation processes by setting a generous softening radius of $r_{\rm soft,ss,2} = 0.01 \Pc$. 
Note that the individual mass of these sinks should be smaller by a factor of several than the 
individual mass of massive stars for the dynamical friction of massive stars to be correctly accounted for.
We refer to the sink particles representing lower mass stars as the second group sinks. 

To take into account the strong dynamical interactions between massive stars, massive stars have 
far smaller softening radii, which correspond to the stellar radius, i.e. $r_{\rm soft,ss,1} = 3\times 10^{-7} \Pc$. 
The sink particles representing the massive stars are referred to as the first group sinks. 

\begin{table*}
\begin{tabular}{ccccccc}
Run name & $M_{\rm cl}$ [$M_{\odot}$] & $M_{\rm gas}$ [$M_{\odot}$] & $m_{\rm up}$ [$M_{\odot}$] & $M_{\rm massive}/M_{\rm cl}$ & $N_{\rm sink,1}$ & $N_{\rm sink,2}$ \\
\hline
% see Exc. book SILCC II from 21st May 2019
C1e3RW &   1000 &  0  &  40 & 0.14 & 689 & 0 \\
C1e3TW &   1000 &  0  &  40 & 0.14 & 9 & 100 \\
\hline
C3e3RG &   1000 &  2000  &  40 & 0.14 & 689 & 0 \\
C3e3TG &   1000 &  2000  &  40 & 0.14 & 9 & 100 \\
\end{tabular}
\caption{Parameters of the simulations investigated in the tests of \refs{sssTwoSoftRadsEC} and \ref{sssTwoSoftRadsGE}. 
The clusters are of mass $M_{\rm cl}$, and they are embedded in a spherical gaseous potential of mass $M_{\rm gas}$. 
The IMF is populated up to mass $m_{\rm up}$, so the total mass of massive stars is $M_{\rm massive}$. 
The clusters are composed of two different groups of sink particles with number 
of stars $N_{\rm sink,1}$ and $N_{\rm sink,2}$ in the respective group. 
}
\label{tListSimsTest}
\end{table*}

\iffigscl
\begin{figure*}
\includegraphics[width=\textwidth]{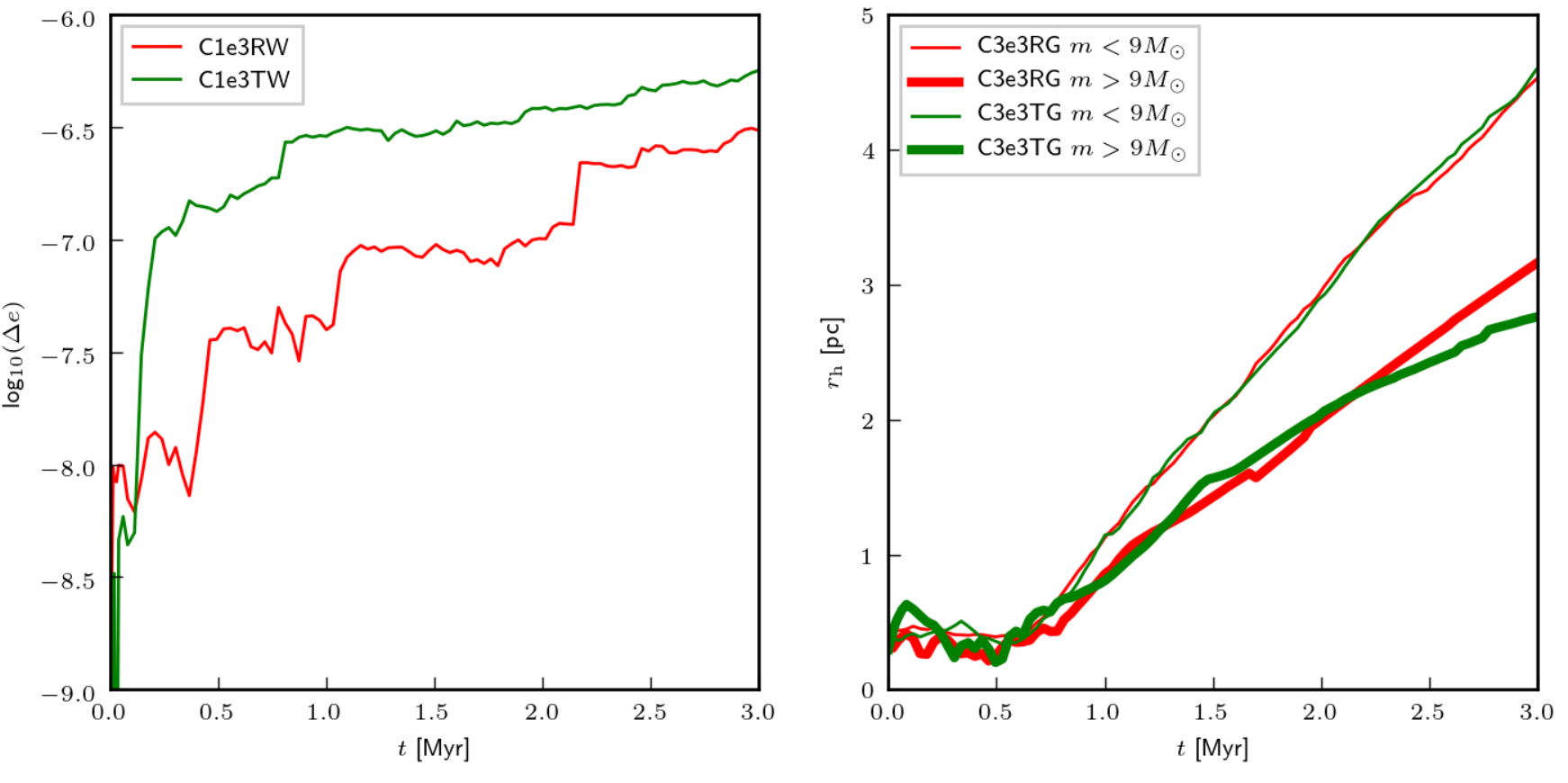}
\caption{\figpan{Left panel:} Evolution of the energy error $\Delta e(t)$ for 
gas free clusters (models of \refs{sssTwoSoftRadsEC}). 
The model with two groups of sink particles (C1e3TW; green line) has only slightly (by $\approx 0.5$ dex) 
larger energy error than the control run of the resolved cluster (C1e3RW; red line). 
\figpan{Right panel:} Evolution of the half-mass radius $r_{\rm h}$ in models C3e3TG (green lines) 
and C3e3RG (red lines), which is studied in \refs{sssTwoSoftRadsGE}. 
The gaseous potential is approximated by an analytic Plummer model, which exponentially decreases 
after $0.6 \Myr$, resulting in rapid expansion of the cluster. 
We separately calculate $r_{\rm h}$ for lower mass stars ($m < 9 \Msun$; thin lines) 
and massive stars ($m > 9 \Msun$; thick lines); $r_{\rm h}$ of both stellar groups 
evolves almost identically between the models.
}
\label{fHalfMass}
\end{figure*} \else \fi

In order to check energy conservation in a cluster with two groups of sink particles, 
we perform the following models. 
Model C1e3TW is the test cluster containing two different groups of sink particles. 
The cluster has a Plummer profile of $a_{\rm Pl} = 0.3 \Pc$, mass of $M_{\rm cl} = 10^3 \Msun$, it contains 
9 massive stars generated from the Kroupa IMF
in the mass range of $(9 \Msun, 40 \Msun)$ 
with total mass in massive stars $M_{\rm massive} = 140 \Msun$. 
These stars are represented by sink particles of group 1.
The lower mass stars are approximated by sink particles of group 2; we adopt in total 
100 sink particles, each of mass $8.6 \Msun$. 

The control model (C1e3RW), which represents a resolved star cluster (i.e. a particle represents a single star), 
has the same $M_{\rm cl}$ and $a_{\rm Pl}$ as model C1e3TW, but
all its stars are sampled from the Kroupa IMF from the range of $(0.5 \Msun, 40 \Msun)$, 
which results in $680$ stars, and 
they all have $r_{\rm soft,ss,1} = 3\times 10^{-7} \Pc$. 
The details of the simulations are summarised in \reft{tListSimsTest}. 

The relative energy error $\Delta e(t)$ between the models is compared in the left panel of \reff{fHalfMass}.
It demonstrates that the approximated model C1e3TW  
with sink particles of two different softening radii has only slightly larger 
energy error (by less than 0.5 dex) than model C1e3RW with all sink particles of identical softening radius. 
Thus, using sink particles of very different softening radii does not introduce substantially 
larger energy error than using sink particles of the same softening radius. 

\subsubsection{Clusters with two different softening radii for sink particles: Reaction on gas expulsion}

\label{sssTwoSoftRadsGE}

The models of the previous section are set up without gas for the purpose to test the mechanical energy conservation. 
However, these models are not that suitable for testing the dynamical evolution of stars. 
For this purpose, we study the half-mass radii evolution of clusters impacted by a rapid change in the gravitational 
potential $\phi$ due to gas expulsion. 
Here, unlike in \refs{sMainSim} below, we prescribe the behaviour of the gas by a simple spherical analytic model, 
so we do not need to introduce gas on the grid. 

Another reason for this test is to verify the approximation of the lower mass stars by the two groups of 
sink particles in a configuration where the gravitational field of gas changes rapidly. 
These conditions are similar to the simulations performed in \refs{sMainSim}, and here we check 
that the massive stars, which feel the gravitational field 
not only from gas but also from lower mass stars, evolve close to the control model.

The gas is modelled by a Plummer model of initial mass $M_{\rm gas}(t=0) = 2 M_{\rm cl}$, and Plummer 
parameter $a_{\rm Pl} = 0.3 \Pc$. 
Gas expulsion is realised by exponentially decreasing the gaseous mass,
\begin{equation}
M_{\rm gas}(t) = M_{\rm gas}(t=0) \exp{\{-(t - t_{\rm d})/\tau_{\rm M}\}}, \; t > t_d
\label{ePotGaseous}
\end{equation}
which is a common approach in N-body simulations (e.g. \citealt{Kroupa2001b,Baumgardt2007} see also \citealt{Goodwin1997} 
for a similar earlier study). 
Gas expulsion starts after time delay of $t_{\rm d} = 0.6 \Myr$, and it occurs on time-scale of $\tau_{\rm M} = 0.03 \Myr$, 
which is shorter than the half mass crossing time of the cluster ($t_{\rm h} = 0.15 \Myr$). 
As in \refs{sssTwoSoftRadsEC}, we run a model with stars represented by two groups of sink 
particles for the lower mass and massive stars 
(model C3e3TG), as well as a model where the same stellar mass is realised by sampling the 
Kroupa IMF in the range of $(0.5 \Msun, 40 \Msun)$ by one group of sink particles (model C3e3RG). 
The simulations are detailed in \reft{tListSimsTest}. 

The right panel of \reff{fHalfMass} shows the time evolution of the half-mass radius $r_{\rm h}$ calculated 
separately for the lower mass stars (thin lines), and the higher mass stars (thick lines).
Before gas expulsion (at $t_d = 0.6 \Myr$), the radius $r_{\rm h}$ remains approximately constant 
for the lower mass stars, while it decreases for more massive stars as these stars mass segregate. 
The fact that $r_{\rm h}$ for lower mass stars stays approximately constant for $0.6 \Myr$, which corresponds to 
several cluster crossing times $t_{\rm h}$, indicates that the second group of sinks is nearly in equilibrium. 

The clusters rapidly expand at $0.6 \Myr$ due to gas expulsion. 
Massive stars, which partially mass segregated, are more centrally concentrated than the lower mass stars. 
The agreement in $r_{\rm h}$ between the approximative model C3e3TG for both groups of sink particles (green lines) 
and model C3e3RG representing a resolved cluster (red lines) is very good. 
The agreement indicates that the adopted approximation to lower mass stars is appropriate for studying 
the reaction of stars (both lower mass and massive) on gas flows in embedded star clusters.

\subsection{Mutual interaction between sink particles and gas}

\label{ssSinksOnGas}

While the main topic of the previous section is to check the accuracy of the integrator 
between sink particles themselves, here we test the accuracy of the integrator between sink particles and gas. 
Note that the main qualitative difference between the Hermite and leap-frog integrators for a particle 
subjected to a gas 
is in the predictor step of the former integrator, which takes into account the velocities of the individual 
tree nodes \eqp{efdot}. 

\iffigscl
\begin{figure*}
\includegraphics[width=\textwidth]{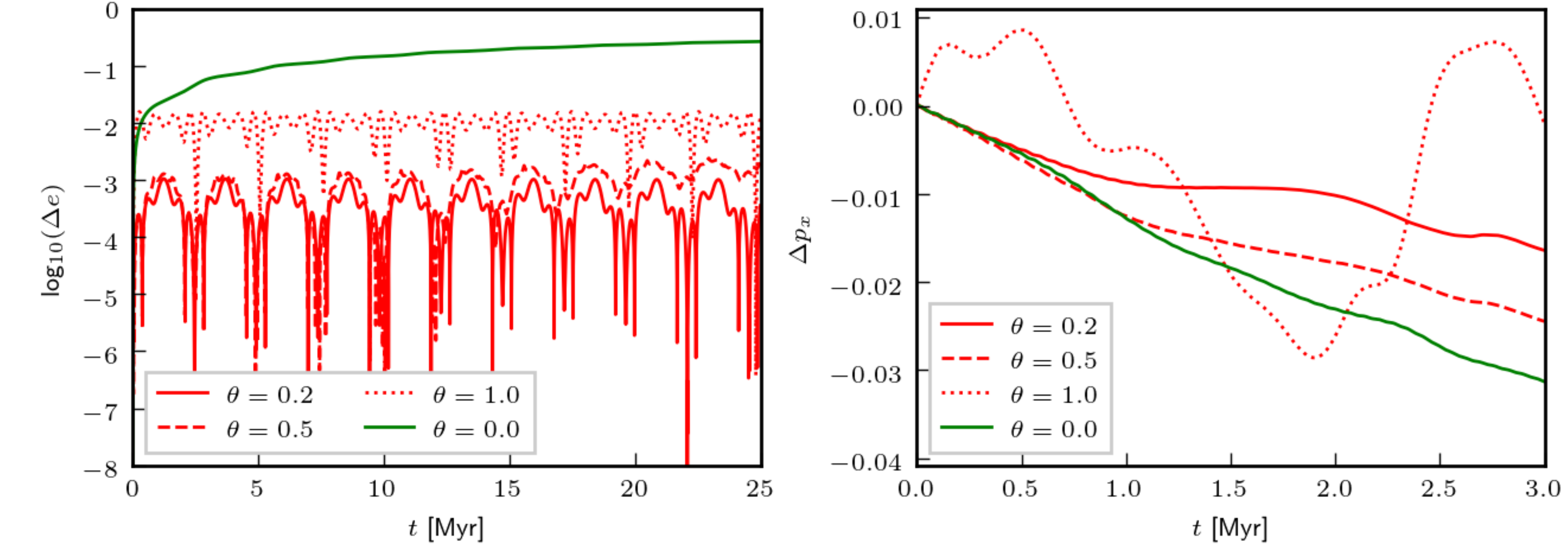}
\caption{\figpan{Left panel:} The energy error evolution $\Delta e(t)$ for a sink particle  
on a circular orbit around a Bonnor-Ebert sphere during ten orbits (\refs{sssGOS}).
The results for the Hermite integrator with different opening angles $\theta$
for blocks are shown by red lines. 
The result for the leap-frog integrator, which uses $\theta=0$, is shown by the green line. 
\figpan{Right panel:} The $x-$ momentum error evolution ($\Delta p_{\rm x}(t) = p_{\rm x}/|\mathbf{p}|$) for 
a system composed of a Bonnor-Ebert sphere and a sink particle (\refs{sssSOG}). 
The line description is the same as in the left panel.
}
\label{fBESphere_2p}
\end{figure*} \else \fi

\subsubsection{A sink particle subjected to gas}

\label{sssGOS}

The initial conditions consist of a stable Bonnor-Ebert sphere of parameter $\xi = 4$, 
mass $M_{BE} = 1 \Msun$, temperature $T = 10\Kk$ and mean molecular weight $\mu = 2.0$. 
The sphere is confined at a radius of $0.052 \Pc$ by a warm medium of temperature $T_{\rm a} = 10^4 \Kk$ with 
mean molecular weight $\mu_a = 0.5$. 
The sphere is placed at the centre of a box of uniform resolution $0.003 \Pc$. 

There is a sink particle of mass $m_{\rm sink} = 0.01 \Msun$ located outside the sphere at 
the position $(x,y,z) = (0.08,0,0) \Pc$ relative to its centre, and 
with initial velocity $v_y = 0.24 \Kms$ pointing in the $y$ direction so that the particle is on a circular trajectory 
around the sphere. 
The trajectory is integrated for $10$ orbits. 

Hydrodynamics is switched off to keep the 
gravitational field as close to the theoretical value as possible. 
The relative energy error 
$\Delta e(t)$ is defined again by \eq{eRadRatio}, but in this test we use $E(t) = v(t)^2/2 - GM_{BE}/r(t)$, 
where $v(t)$ and $r(t)$ is the instantaneous sink particle velocity and position relative to the centre of the 
Bonnor-Ebert sphere. 
Note that this approach neglects the mass of the low density warm gas (of density by $\sim 10^3$ lower than the density 
of the sphere) outside the Bonnor-Ebert sphere, which likely increases the energy error. 
To test the influence of the discretisation of the Bonnor-Ebert sphere to tree nodes, 
we perform the calculation with three values of the opening angle $\theta$ \citep{Barnes1986}.
The left panel of Figure \ref{fBESphere_2p} shows the evolution of the relative energy error $\Delta e(t)$ 
for the Hermite integrator (red lines) with $\theta = 0.2$, $\theta = 0.5$ and $\theta = 1$. 
As expected, the energy error decreases with decreasing $\theta$; from $\approx 10^{-2}$ 
for $\theta = 1$ to $\approx 10^{-3.5}$ for $\theta = 0.2$.

The energy error rather periodically fluctuates than increases. 
The fluctuations are likely caused by the finite discretisation of the Bonnor-Ebert sphere on the grid, which 
introduces a non-spherical perturbations. 
However, perturbations are symmetric to planes $x = 0$, $y = 0$, and $z = 0$, which implies that 
their contributions cancel out as the sink particle orbits (for example, the perturbation to 
the force component $F_x$ at $(x,y,0)$ is of the same value but of the opposite sign than that at $(-x,y,0)$). 
This explains why the energy error suddenly drops when the particle passes through its initial position, 
which happens every $\approx 2.5 \Myr$. 
This means that the already low energy error is dominated by the finite resolution of the Bonnor-Ebert sphere 
instead of the properties of the integration scheme. 
Even lower energy error will be obtained for higher resolution simulations.

In contrast, the leap-frog integrator has substantially larger energy error, which 
reaches $10^{-1}$ after several orbits (left panel of \reff{fBESphere_2p}).  
This is despite the fact that the leap-frog integrator evaluates the gravitational force at a higher accuracy 
(and at higher CPU costs), 
resolving contribution from each cell individually ($\theta = 0$). 
The Hermite integrator outperforms the leap-frog even with a generous resolution of the blocks for gravity 
with $\theta = 1$ as seen by the red dotted line.

\subsubsection{Gas subjected to a sink particle}

\label{sssSOG}

The initial conditions of this test are identical to the test in \refs{sssGOS}, with the only difference 
that the hydrodynamics is enabled. 
This unfortunately makes the test less sensitive because the gravitational field is no longer spherically-symmetric, 
but such a high accuracy is not necessary for the present test. 

As the particle orbits the Bonnor-Ebert sphere, the sphere accelerates towards the particle, and also 
the particle stirs tides on the surface of the sphere. 
However, the total momentum of the system should conserve, which we investigate here. 
Because the sphere is at rest initially, the total momentum corresponds to the initial momentum of the sink particle 
$\mathbf{p} = (0, m_{\rm sink} v_y, 0)$. 
The right panel of \reff{fBESphere_2p} plots the evolution of the relative error of the 
conservation of the $x-$ momentum component,  
i.e. $\Delta p_{x}(t) \equiv p_x/|\mathbf{p}|$, for the Hermite integrator (red lines) 
with $\theta = 0.2$, $\theta = 0.5$ and $\theta = 1$, and for the leap-frog integrator (the green line). 
The plotted time-scale is slightly longer than one particle orbit. 
For the Hermite scheme, the momentum error decreases with decreasing $\theta$ from 3\% for $\theta=1.0$ to 
1.5\% for $\theta=0.2$. 

The Hermite integrator again outperforms the leap-frog integrator (green line) 
despite leap-frog having the entering gravitational force determined with 
higher accuracy (leap-frog has $\theta=0$, while Hermite integrator has $\theta > 0$). 
This also results in faster code execution of the Hermite scheme because it needs a smaller number of tree nodes 
to be opened for the same accuracy.

\subsubsection{Gravitational collapse and accretion}

The main purpose of this test is to check the accuracy of the gravitational force which 
drives the infall of the outer parts of the sphere towards the centre, where it gets 
accreted onto the sink particle.
The force is generated by both the innermost gas and the sink particle, and as the 
sink particle grows in mass, its gravity becomes more important while the accretion rate should remain constant until 
the sphere is exhausted. 
Thus, if the acceleration due to gas differed from the acceleration due to the sink, 
the accretion rate would change as the gravity from the sink particle becomes more significant with time. 
This configuration is suited for testing possible imbalances between the gravitational force 
due to sink particles and gas as experienced by a gaseous body.  

As derived by \citet{Shu1977}, a core surrounded by a spherically symmetric isothermal gas with density in the form of
\begin{equation}
\rho(r) = \frac{A c_s^2}{4 \pi G r^2},
\label{eIsothermalProfile}
\end{equation}
accretes the gas at the rate of
\begin{equation}
\dot{M} = \frac{m_0 c_s^3}{G},
\label{eAccRate}
\end{equation}
where $c_s$ is the sound speed, and $m_0$ and $A$ dimensionless constants. 
The constant $m_0$ is a function of $A$. 

We initialised the sphere with $c_s = 0.2033 \Kms$ and $A = 29.3$. 
This value of $A$ results in $m_0 = 133$ \citep{Federrath2010}, so 
the theoretical accretion rate \eqp{eAccRate} is $265 \Msun/\Myr$. 
We study the influence of increasing spatial resolution as well as the accuracy of gravitational force evaluation 
(controlled by the opening angle $\theta$). 
For both the Hermite and leap-frog integrator, we run three models of increasing spacial resolution 
from $\Delta x = 0.063 \Pc$, via $0.031 \Pc$ to $0.016 \Pc$ and decreasing $\theta$ from 
$\theta = 1.0$, via $0.5$ to $0.2$ (e.g. the model with $\Delta x = 0.016 \Pc$ has $\theta = 0.2$). 
We place a sink particle of negligible mass $1 \rm{g}$ at rest at the centre of the sphere so that 
we can measure its accretion rate, which we use as an indicator of the accuracy by which the 
gravitational force of the sink particle and the gas attracts the gas from the outer parts of the sphere. 
We set both the accretion radius $r_a$ and the softening radius $r_{\rm soft,sg}$
for gas-sink interaction to $2.5 \Delta x$. 
The threshold density for accretion $\rho_{\rm thr}$ is determined as $\rho_{\rm thr} = \pi c_s^2/(4 G r_a^2)$. 

\iffigscl
\begin{figure}
\includegraphics[width=\columnwidth]{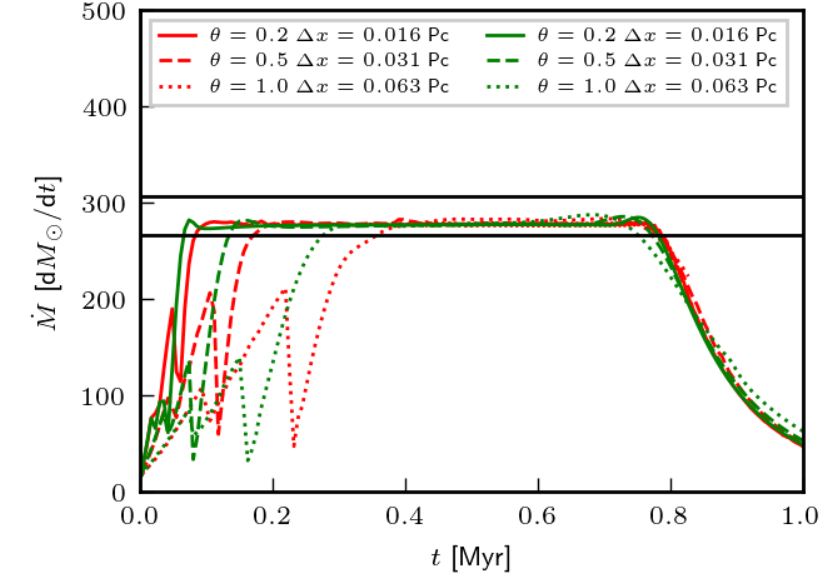}
\caption{The accretion rate onto an isothermal sphere calculated by the \textsc{HermiteSink} module 
(red lines) as compared to the standard sink particle implementation in \flashd (green lines). 
For each of the implementation, we vary the opening angle $\theta$ as well as spacial resolution $\Delta x$ 
as indicated by the line style. 
The theoretical accretion rates for temperature $10 \Kk$ and $11 \Kk$ according to \eq{eAccRate} 
are indicated by the horizontal black lines (the initial temperature of $10 \Kk$ increases 
due to the large compression, the ratio of specific heats is $\gamma = 1.001$).}
\label{fAccRate}
\end{figure} \else \fi

\reff{fAccRate} compares the evolution of the accretion rate as calculated by the \textsc{HermiteSink} module 
(red lines) and by the  standard \flash sink particle implementation (green lines). 
For both integrators, higher resolution models reach the highest accretion rate earlier. 
After establishing, the accretion rate is almost constant in all the models, and the value of this 
constant does not depend on the resolution. 
The accretion rate calculated by the \textsc{HermiteSink} module is almost identical to the 
standard \flash sink particle implementation, and in agreement with \eq{eAccRate} as indicated by the 
horizontal lines. 

%\subsubsection{Boss \& Bodenheimer test}

%Once with sinks in tree and without tree on another time

\subsection{Practical aspects and limitations}

\label{ssPractical}

The integration scheme is coupled to the stellar evolutionary module as well as the modules dealing with 
stellar winds, ionising radiation and SNe, which are described in \citet{Gatto2017} and \citet{Peters2017}.
This enables self-consistent treatment of stellar dynamics and feedback. 

To provide an order of magnitude estimate of the CPU costs on a modern processor (Xeon E5-2697 v3), 
integrating a cluster of $10^3$, $3 \times 10^3$ and $10^4$ stars for 10 $r_{\rm h}$ crossing times lasts for $1.2 \times 10^2$s, 
$3.1 \times 10^3$s, and $10^5$s, respectively (\reff{fClusters}). 
In these cases, we adopt rather stringent criteria (compact cluster of $a_{\rm Pl} = 0.1 \Pc$, 
small softening radius of $r_{\rm soft,ss} = 5 \times 10^{-6} \Pc$ and stars of different masses) 
to obtain the likely upper limit of CPU demands in realistic simulations. 
Likewise, if the same number of stars is distributed to more clusters, the calculation is substantially faster. 
Since the N-body integration is done in serial, the integrator simply adds these CPU demands to the total costs. 
Taking into account the huge CPU costs typically consumed by modelling other physical processes (e.g. gas self-gravity, radiative 
transfer) in state-of-the-art high resolution simulations, 
integration of a compact cluster with up to $N = 10^4$ stars is entirely possible with the \textsc{HermiteSink} module.
Possibly larger number of sink particles can be integrated in more rarefied clusters or with using 
a larger softening radius.

\section{The timescale of gas expulsion from embedded star clusters}

\label{sMainSim}

%Generate a spherically symmetric cloud with cluster at the centre.
%
%Answer one of the following two questions:
%
%1. What is the timescale of gas removal? - in this paper
%
%2. Which fraction of gas can be removed as a function of the escape speed from the cluster (i.e. varying cluster mass)? 
%Is there a threshold of $10 \Kms$?

\subsection{Numerical method and initial conditions}

\label{ssGEXInit}

\subsubsection{Numerical method}

The simulations are performed by 3D adaptive mesh refinement code \flash \citep{Fryxell2000}. 
Hydrodynamics is advanced by the PPM method \citep{Colella1984}. 
Self-gravity 
due to gas is evaluated by an octal tree method with geometric criterion and opening angle $\theta = 0.5$ \citep{Wunsch2018}. 
Sink particles are integrated by the \textsc{HermiteSink} module described above in this paper. 
We use two groups of sink particles, where the first group sinks (representing massive stars) 
are sources of ionising radiation and stellar winds, 
while the second group sinks (lower mass stars) interact only gravitationally. 
The number of ionising photons as well as the properties of stellar winds are 
obtained from synthetic stellar evolutionary tracks of \citet{Ekstrom2012}. 
These quantities are calculated for each star individually based on its mass $m$ and age. 
The transfer of ionising radiation is calculated by the \textsc{TreeRay} algorithm (W\"{u}nsch et al. in preparation). 
Stellar winds are realised by injecting the wind momentum to a sphere of the radius of three grid cell size 
around the star (which is at the highest refinement level), which is described in detail in \citet{Haid2018}. 

The interstellar medium (ISM) is represented by a two fluid model, 
where the cold phase is isothermal and the warm and hot phases are 
adiabatic with polytropic index $\gamma = 5/3$. 
The cold isothermal phase is always at $10 \Kk$, and it turns into the other fluid if 
irradiated by ionising radiation from massive stars (temperature $10^4 \Kk$). 
Then, it can be heated further by thermalisation due to stellar winds. 
The gas at temperature in excess of $10^4 \Kk$ cools according to the \citet{Sutherland1993} cooling function. 
The rapid cooling in the warm and cold phase is approximated by setting the 
temperature immediately to $10 \Kk$ for any non-irradiated cell which reached temperature $10^4 \Kk$.

\subsubsection{Initial conditions}

\begin{table*}
\begin{tabular}{cccccccccccc}
\multicolumn{2}{c}{Run name} & $M_{\rm embd}$ & $M_{\rm cl}$ & $M_{\rm gas}$ & $m_{\rm up}$ & $N_{\rm massive}$ & $\frac{M_{\rm massive}}{M_{\rm cl}}$ & $t_{\rm ff}$ &  
$\overline{\Sigma}_{\rm gas}$ & $R_{\rm str}$ \\
 & & [$10^3 M_{\odot}$] & [$10^3 M_{\odot}$] & [$10^3 M_{\odot}$] & [$M_{\odot}$] & & & [$\Myr$] & [$\Sd$] & [$\Pc$] \\
\hline
% see Exc. book SILCC II from 21st May 2019
% calculated by script cluster_properties.py located in /home/franta/Desktop/git/kratke_vypocty/cluster_properties
    C9e2 &     O9e2 &  0.9  &  0.3 &  0.6   &       19 &    2 & 0.08 &   0.96 &      96 &     0.07 \\
    C3e3 &     O3e3 &  3.0  &  1.0 &  2.0   &       40 &    7 & 0.14 &   0.52 &     320 &     0.11 \\
    C9e3 &     O9e3 &  9.0  &  3.0 &  6.0   &       71 &   29 & 0.17 &   0.30 &     960 &     0.11 \\
    C3e4 &     O3e4 &  30.0 &  9.0 & 20.0   &      112 &   90 & 0.18 &   0.17 &    3200 &     0.09 \\
    C9e4 &     O9e4 &  90.0 & 30.0 & 60.0   &      120 &  276 & 0.20 &   0.10 &    9600 &     0.06
\end{tabular}
\caption{List of the modelled embedded clusters. First two columns are model names, 
$M_{\rm embd}$ is the total mass of the embedded cluster (i.e. stellar and gaseous mass combined), 
$M_{\rm cl}$ is the total stellar mass of the cluster (both in low mass and massive stars), 
$M_{\rm gas}$ is the initial gaseous mass, 
$m_{\rm up}$ is the upper mass limit of stars in the cluster, 
$N_{\rm massive}$ is the total number of massive stars,
$M_{\rm massive}$ is the total mass in massive stars, 
$t_{\rm ff}$ is the initial central free fall time, 
$\overline{\Sigma}_{\rm gas} (1 \Pc)$ is the mean gas surface density within projected radius of $1 \Pc$, 
and $R_{\rm str}$ is the initial Stromgren radius of all stars if they were located at one point at the cluster centre.}
\label{tListSimsMain}
\end{table*}

The embedded clusters consist of two components: a gaseous component (of total initial mass $M_{\rm gas}$), 
and stellar component (of total initial mass $M_{\rm cl}$), 
so the total embedded cluster mass is $M_{\rm embd} = M_{\rm cl} + M_{\rm gas}$. 
The most important properties of the simulations are summarised in \reft{tListSimsMain}.
In all the simulations, we set the star formation efficiency $\sfe = M_{\rm cl}/(M_{\rm cl} + M_{\rm gas})$ 
to $1/3$ regardless of the cluster mass. 
Although higher SFEs are suggested by some works (particularly for more massive clusters, e.g. 
\citealt{Kruijssen2012}), another works suggest the value of SFE to be around $1/3$ \citep[e.g.][]{Lada2003,Banerjee2017}, 
so the value of the SFE is not well constrained.  
We adopt a relatively low value of the SFE to study 
the upper limit on the gas expulsion time-scale as well as the maximum cluster mass for which  
the cloud dissolution by photoionising and wind feedback is inevitable. 

The stellar as well as gaseous component is represented by a Plummer sphere of Plummer parameter $a_{\rm Pl} = 1 \Pc$. 
This initial radius is somewhat larger than the majority of observed embedded star 
clusters (e.g. \citealt[][]{Kuhn2014,Traficante2015} see also \citealt{Marks2012}). 
The main reason for our choice of the value of $a_{\rm Pl}$ is to avoid clouds of surface density $\Sigma_{\rm gas}$ 
in excess of $\sim 5000 \Msun$ (c.f. \reft{tListSimsMain}), 
which have feedback dominated by radiation pressure \citep{Krumholz2009,Fall2010}, 
which we cannot model by \flash currently. 
Another reason is that the cloud collapses at the beginning of the simulation, so the 
stellar component, which follows the gravity of the gas, becomes more concentrated before feedback (in some 
of the models) reverses the collapse. 

The embedded cluster is placed at the centre of a cube of side-length $10 \Pc$ 
with a basic resolution grid of $\Delta x = 0.08 \Pc$. 
The grid is adaptively refined on sink particles up to the highest refinement level with resolution of $\Delta x = 0.02 \Pc$. 
This resolution enables us to resolve the Stromgren radius even at the centre of the cloud if all 
massive stars are located at the centre (c.f. $R_{\rm str}$ in \reft{tListSimsMain}).
We do not base the grid refinement criterion on the Jeans length because we are not interested in capturing 
the details of gravitational collapse; instead, we intend to resolve the immediate vicinity of stars because 
it is where feedback is imparted to the gas.

The star cluster is initially in virial equilibrium with the gas; the initial sink particle positions and velocities 
are generated by the method of \citet{Aarseth1974b}. 
% In order to save CPU time, the star clusters are populated by stars as follows. 
Since feedback produced by a star depends sensitively on its mass, it is 
necessary to limit the upper stellar mass $m_{\rm up}$ for the lower mass clusters; for example, generating a $120 \Msun$ star 
in a $1000 \Msun$ cluster would overestimate destruction of its natal cloud, yet such massive stars 
are rarely observed in lower mass clusters. 
We adopt the particular value of $m_{\rm up}$ for a cluster 
of mass $M_{\rm cl}$ from the $m_{\rm up} - M_{\rm cl}$ relation proposed 
by \citet{Weidner2010} (see also \citet{Elmegreen1983,Elmegreen2000,Weidner2004,Oey2005} for earlier works). 
Accordingly, massive stars ($m > 9 \Msun$; represented by the first group of sink particles) 
are drawn randomly from the IMF of \citet{Kroupa2001a} from 
the mass interval $(9 \Msun, m_{\rm up})$. 
This means that with increasing cluster mass $M_{\rm cl}$, the total mass in massive stars $M_{\rm massive}$ 
increases not only because of larger cluster mass, but also because the IMF is populated towards more massive stars 
(c.f. the ratio $M_{\rm massive}/M_{\rm cl}$ in \reft{tListSimsMain}). 

This sampling of the IMF has the interesting consequence that 
the strength of feedback, as measured in the number of ionising photons $N_{\rm phot}$ 
and the stellar wind mechanical luminosity $L_{\rm wind}$, increases more than linearly with cluster mass 
as can be seen in \reff{ffeedback}. 
Moreover, both $N_{\rm phot}$ and $L_{\rm wind}$ 
increase even steeper than $M_{\rm cl}^2$ in the cluster mass interval $M_{\rm cl} = (10^{2.5} \Msun, 10^{3.2} \Msun)$, 
which is steeper than the gravitational binding energy (which scales as $M_{\rm cl}^2$). 
The well known argument that gravity wins over feedback above certain mass $M_{\rm cl}$ does not hold in this 
mass range (it holds for clusters with $M_{\rm cl} \gtrsim 10^4 \Msun$ because of saturating $m_{\rm up}$). 

We assume that all stars are already formed at the beginning of the simulation, 
and we disable further formation of sink particles. 
The stellar dynamics of massive stars is calculated accurately (i.e. without the softening approximation)  
by setting the softening radii of massive stars to the stellar radius of the B0 star, 
i.e. $r_{\rm soft,ss,1} = 3 \times 10^{-7} \Pc$.

% the figure is produced by script feedback.py located in /home/franta/Desktop/git/kratke_vypocty/stellar_evolution/feedback/silcc_ekstrom_2012
\iffigscl
\begin{figure}
\includegraphics[width=\columnwidth]{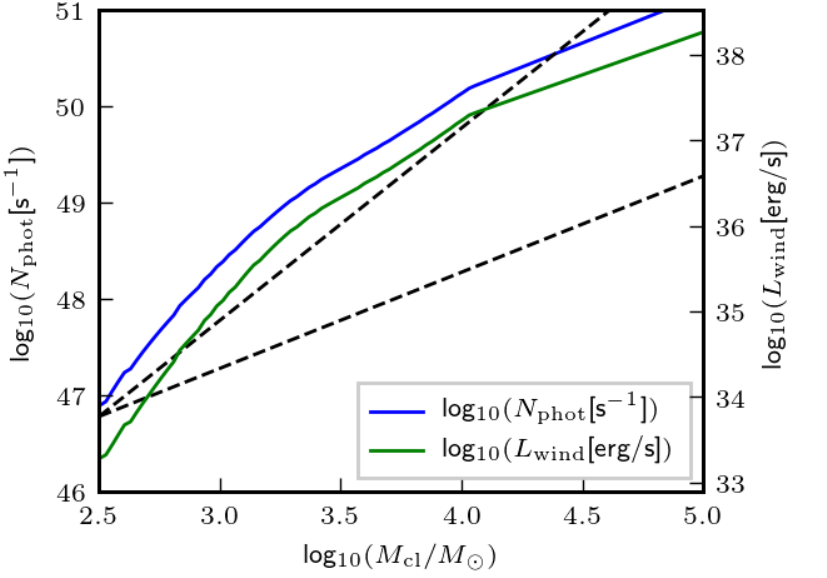}
\caption{The number of hydrogen ionising photons $N_{\rm phot}$ (blue line) and the stellar wind mechanical 
luminosity $L_{\rm wind}$ (green line) as a function of the cluster mass $M_{\rm cl}$. 
The stellar IMF was sampled randomly from the IMF of \citet{Kroupa2001a} assuming 
the $m_{\rm up} - M_{\rm cl}$ relation of \citet{Weidner2010}. 
The dashed lines indicate slopes $\propto M_{\rm cl}$ and $\propto M_{\rm cl}^2$. 
With the adopted IMF sampling, $N_{\rm phot}$ and $L_{\rm wind}$ increase steeper than linearly 
in the mass interval $M_{\rm cl} = (10^{2.5} \Msun, 10^{4} \Msun)$, and even steeper than $M_{\rm cl}^2$
in the mass interval $M_{\rm cl} = (10^{2.5} \Msun, 10^{3.2} \Msun)$.}
\label{ffeedback}
\end{figure} \else \fi

The lower mass stars, which dominate the cluster mass, 
are represented by $100$ sink particles (second group sink particles) each of the same mass. 
This means that a second group sink particle has a mass of $(M_{\rm cl} - M_{\rm massive})/100$. 
The sink particles of both groups follow the Plummer model with the same parameter $a_{\rm pl}$, 
so the clusters are not mass segregated. 
The softening radius is $r_{\rm soft,ss,2} = 0.06 \Pc = 3 \Delta x$. 
We set $r_{\rm soft,sg} = 0.06 \Pc$ for both groups of sink particles. 

The physical reason for this approximation to the lower mass stars is to capture the 
expansion and possible deformation of the cluster due to the change of the gravitational potential 
of the gaseous component, which gravitationally collapses and is pushed by the stellar feedback at the same time. 
The lower mass stars represented by the second sink group, which dominate the stellar mass of the cluster 
(they constitute more than $80$\% of the cluster stellar mass), 
are subjected to the gravitational force, expand, and by this process shallow the potential 
in which the massive stars move. 
This approach has important advantages over their representation by an analytical potential, because an analytical 
potential cannot react on the changing gravitational field generated by the gas, 
and an analytic potential is also usually spherically symmetric while the second group of sinks 
can form a complicated 3D structure. 
The large softening radius $r_{\rm soft,ss,2}$ prevents the second group sink particles from changing their velocity 
substantially during encounters, slowing the process of relaxation, and making the system closer to collisionless. 
% Their number of $100$ make their integration cheap. 
What this approximation neglects is the slow evaporation of the least massive stars, but this 
process typically does not play a substantial role on the cluster dynamics on time-scale of several Myr.
The dynamical suitability of this approximation was verified 
in a very similar model C3e3TG (c.f. \refs{sssTwoSoftRadsGE}), 
where gas was represented by an analytical potential. 

For each embedded cluster mass, we perform two models. 
%The gaseous distribution is identical in both models. 
The first model (models starting with letter "C") features a resolved star cluster 
with massive stars distributed over its volume. 
The other model (models starting with letter "O") is identical to model "C" of the same mass with the only exception 
that the massive stars are located at one point at the centre of the gaseous Plummer sphere. 
To have the same feedback strength in both models, we take the same massive stars from model "C" and move them to 
the centre to produce models "O" with one source of feedback. 
The "O" models might be viewed as models with extreme degree of mass segregation (yet neglecting the 
strong dynamical interactions between the massive stars packed into such a small volume), 
while "C" models are not mass segregated. 
The number after "C" or "O" represents the total mass $M_{\rm embd}$ of the embedded star cluster.
Since otherwise the models have the same properties, we list them at the same line in \reft{tListSimsMain}.

The calculations are terminated at $1 \Myr$, at which time the most of the clouds are either 
dispersed due to feedback, or collapsed to a dense compact structure whose further realistic calculation 
would necessitate star formation, which is disabled in present simulations.

\subsection{Results}

\subsubsection{Overall evolution of embedded clusters}

\iffigscl
\begin{figure*}
\includegraphics[width=\textwidth]{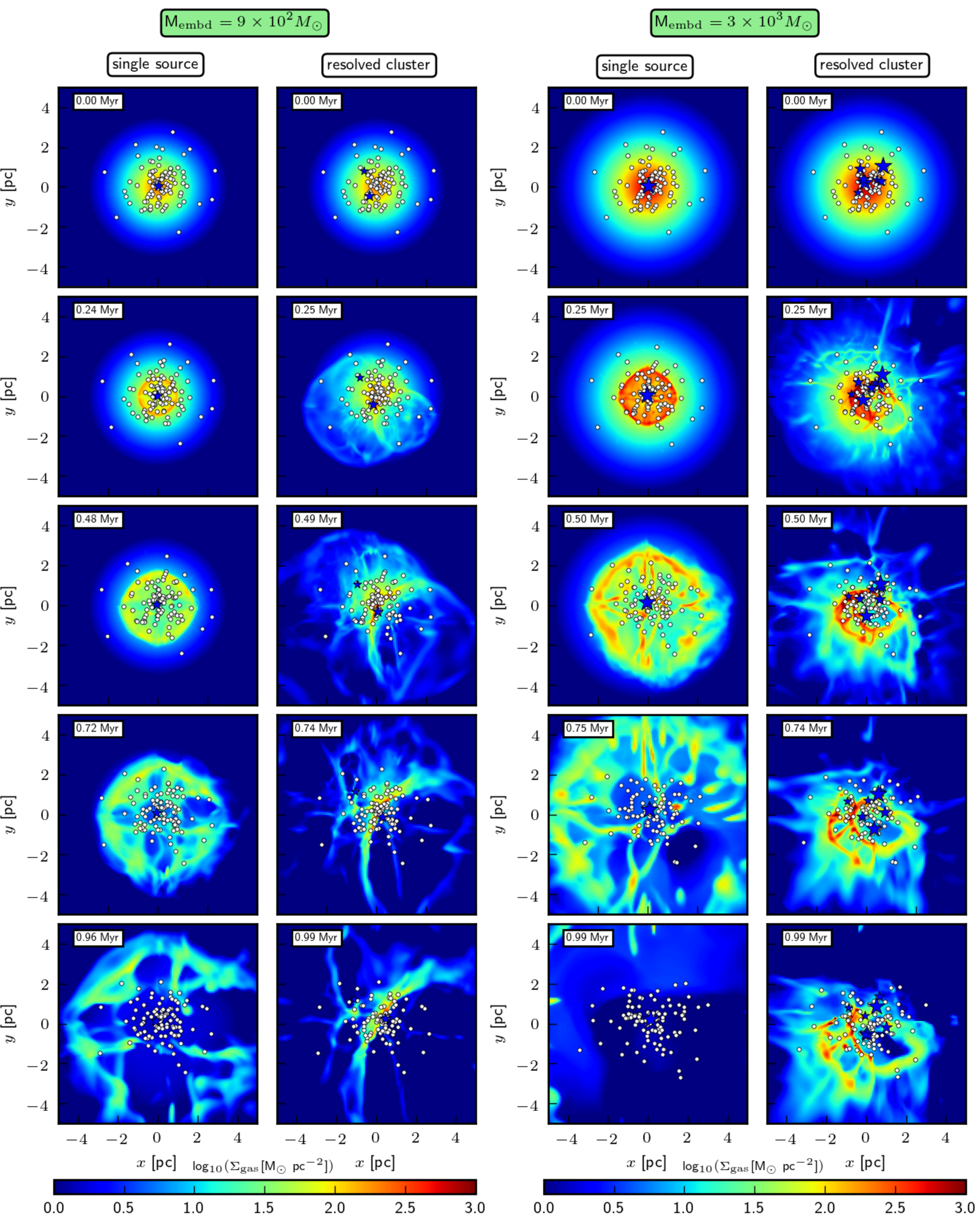}
\caption{Gas column density $\Sigma_{\rm gas}$ evolution for models O9e2 (left column), C9e2 (centre left column), 
O3e3 (centre right column) and C3e3 (right column). 
Massive stars are shown by blue asterisks of the size corresponding to their mass; 
the sink particles representing the lower mass stars are shown by white circles. 
Time (in Myr) is indicated at the upper left corner of each panel. 
Thus, the two left columns compare the evolution for clusters of $M_{\rm embd} = 900 \Msun$ 
for the two extreme distributions of massive stars within the clouds (concentrated at the centre and 
spatially distributed), and 
the two right columns do the same for clusters of $M_{\rm embd} = 3000 \Msun$.
}
\label{fse01}
\end{figure*} \else \fi

\iffigscl
\begin{figure*}
\includegraphics[width=\textwidth]{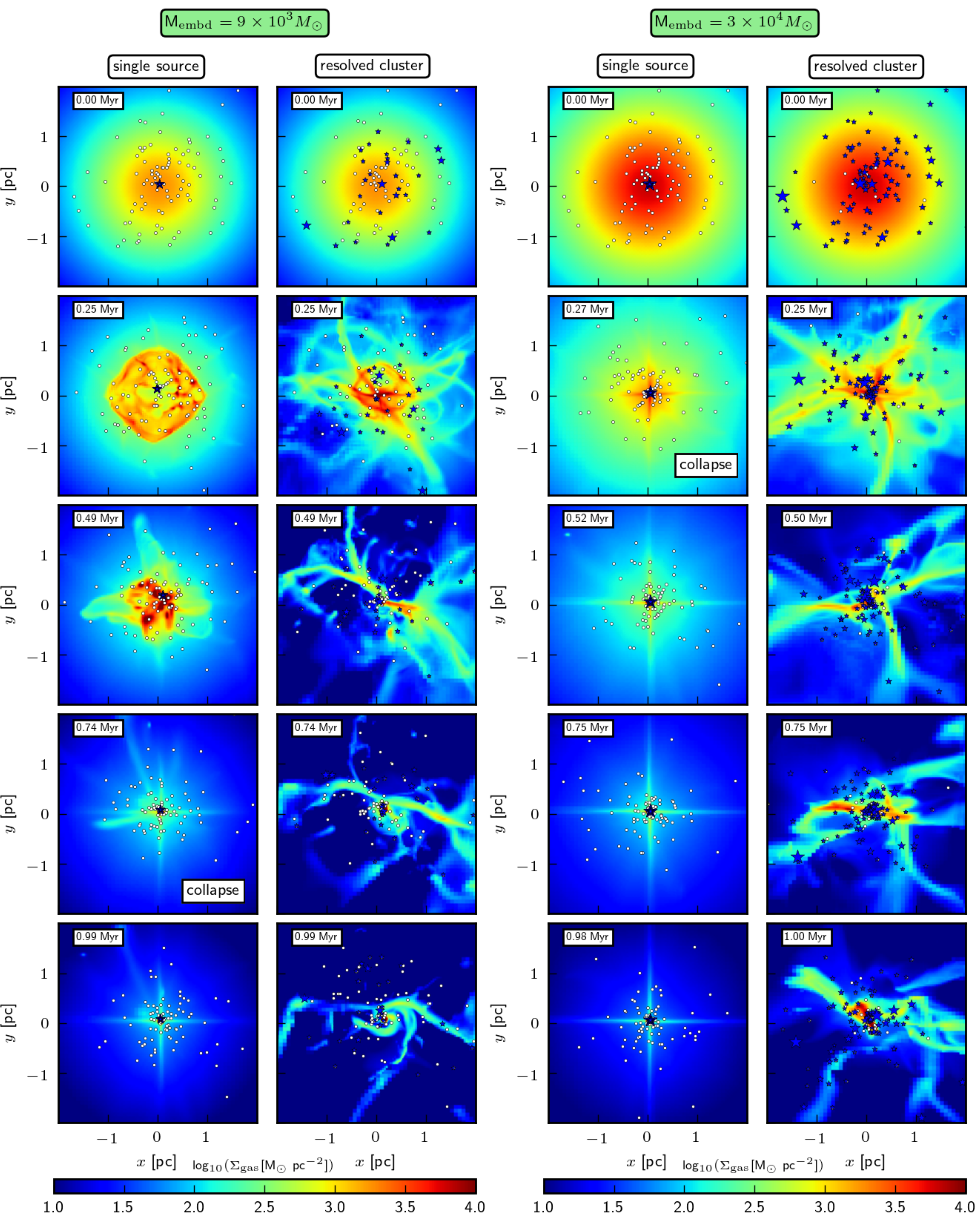}
\caption{The same as \reff{fse01} for models (from left to right) O9e3, C9e3, O3e4 and C3e4, but
note that the scale of $\Sigma_{\rm gas}$ as well as the star size is different than in \reff{fse01}. 
Models O9e3 and O3e4 collapsed and quenched the source by $0.74 \Myr$ and $0.27 \Myr$, respectively.
}
\label{fse23}
\end{figure*} \else \fi

\iffigscl
\begin{figure}
\includegraphics[width=\columnwidth]{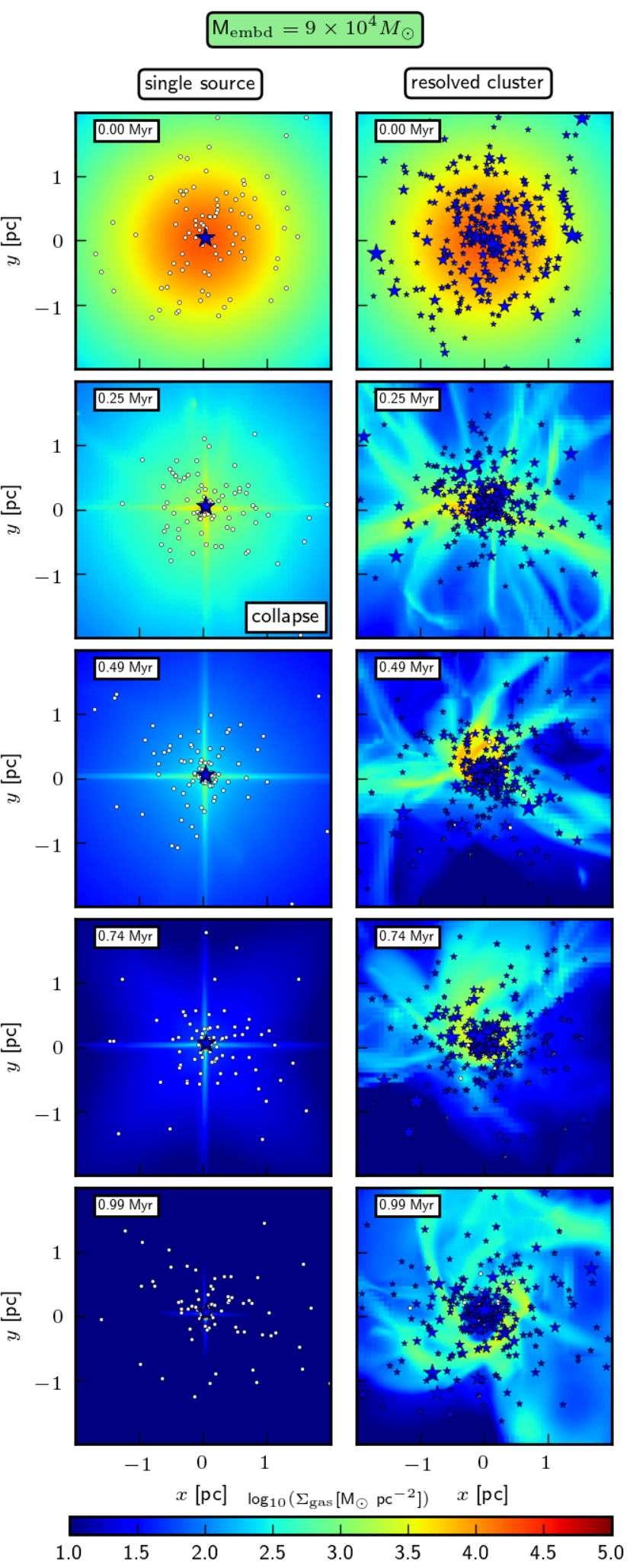}
\caption{The same as \reff{fse01} for models (from left to right) O9e4 and C9e4, 
but note that the scale of $\Sigma_{\rm gas}$ as well as the star size is different than in \reff{fse01}.
Model O9e4 collapsed and quenched the source early by $0.25 \Myr$.
}
\label{fse4}
\end{figure} \else \fi

\iffigscl
\begin{figure*}
\includegraphics[width=\textwidth]{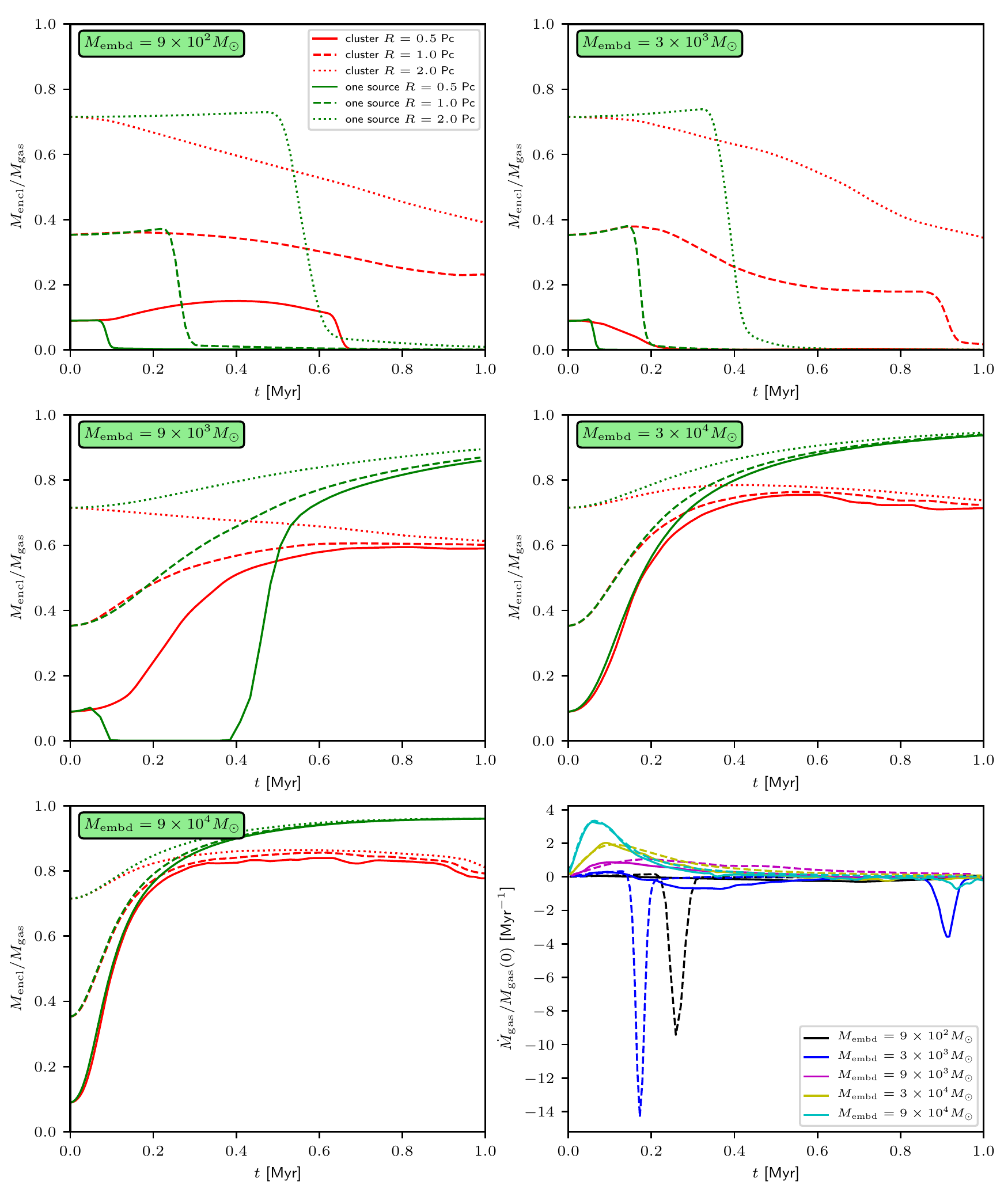}
\caption{\figpan{All panels apart from lower right:} The evolution of the total gaseous mass within a radius of $0.5 \Pc$ (solid lines), 
$1.0 \Pc$ (dashed lines) and $2.0 \Pc$ (dotted lines) for resolved star clusters ("C" models; red lines) 
and for the models with all stars located at one point at the centre ("O" models; green lines). 
The mass of the star cluster is indicated at the upper left of each panel. 
Lower mass clusters ($M_{\rm embd} \leq 3\times 10^3 \Msun$) have substantially larger gas expulsion time-scale $t_{\rm ge}$ 
when the cluster is resolved to individual stars ("C" models). 
Higher mass clusters ($M_{\rm embd} \geq 9\times 10^3 \Msun$) are not able to stop the gaseous infall; 
however resolved clusters are able to reverse the inflow partially and expel $10$ \% to $30$ \% of the gaseous 
mass depending on $M_{\rm embd}$.
\figpan{Lower right:} The mass flow rates through the sphere of radius $1 \Pc$. 
$\dot{M}_{\rm gas} > 0$ indicates gas inflow to the sphere. 
For each mass, models "C" and "O" are represented by the solid and dashed lines, respectively.
}
\label{fgasExp}
\end{figure*} \else \fi

The dynamical evolution of the gaseous and stellar component for each of the embedded cluster 
is shown in Figs. \ref{fse01} through \ref{fse4}. 
Model of given mass $M_{\rm embd}$ is represented by two columns, 
which compare the state of 
the idealised cluster model where all massive stars are located at one point in its centre ("O" models; left column) 
with the resolved cluster ("C" models; right column). 
In addition, we provide movies for the density and temperature evolution for models of $M_{\rm embd} = 3 \times 10^{3} \Msun$ 
and $M_{\rm embd} = 3 \times 10^{4} \Msun$ in the online material. 

The idealised lower mass clusters (models O9e2 and O3e3; \reff{fse01}) drive almost spherical 
shell with origin at the cluster centre, which sweeps up the cloud as it expands. 
The shell accelerates and breaks upon reaching the rarefied outer parts of the cloud 
(at $0.6$ to $0.8 \Myr$), leaving behind an exposed star cluster. 
In contrast, their resolved counterparts (models C9e2 and C3e3) do not drive a single shell, 
but they produce several smaller shells, which interact in a complicated way forming 
open cavities with champagne flows (for example in model C9e2 at $0.49 \Myr$ at the lower right of the panel) as well 
as walls and filaments at their intersections. 
Dense gas then flows along these structures either inwards or outwards depending on the 
interplay between feedback and gravity at the particular place. 
From the figures, it follows that it is more difficult for feedback in the "C" models to push 
the residual gas out of the cluster. 

In order to measure gas expulsion quantitatively, we plot the gas mass $M_{\rm encl}$ enclosed 
inside spheres centred at $(x,y,z) = (0,0,0)$ of a given radius $r_{\rm encl}$ (\reff{fgasExp}).
We consider the radius of $0.5 \Pc$ (solid lines), $1.0 \Pc$ (dashed lines) and $2.0 \Pc$ (dotted lines). 
%The idealised one source models ("O" models) are shown by green lines; the resolved clusters ("C" models) 
%are shown by red lines. 
The distinct shells sweeping the one source models ("O" models; green lines) cause 
a sudden drop of the enclosed mass when they pass through given $r_{\rm encl}$. 
For example in model O9e2, the shell reaches $r_{\rm encl} = 0.5 \Pc$ at $0.08 \Myr$, 
$r_{\rm encl} = 1.0 \Pc$ at $0.25 \Myr$, and $r_{\rm encl} = 2.0 \Pc$ at $0.6 \Myr$. 
The inner part of the shell is substantially rarefied as compared to the initial state of the cloud. 
Before the shell sweeps the material, the mass $M_{\rm encl}$ is almost unaffected, which indicates that 
sweeping by a shell is the dominant process in excavating the clusters. 

In contrast, the resolved clusters ("C" models; red lines) show more complex behaviour of the enclosed mass $M_{\rm encl}$ 
with outflows through some radii and inflows through another. 
Taking model C3e3 as an example, the inner ($r_{\rm encl} = 0.5 \Pc$) and outer ($r_{\rm encl} = 2.0 \Pc$) radii 
have decreasing $M_{\rm encl}$ from the beginning. 
The decrease of $M_{\rm encl}$ at the inner radius is caused by several asymmetric shells 
expanding near the cluster centre, while the decrease of $M_{\rm encl}$ at the outer radius 
is caused by photoerosion due to a star located outside the cluster, which drives an inward propagating 
ionisation front with a champagne flow \citep{TenorioTagle1979,Whitworth1979} streaming outwards. 
The value of $M_{\rm encl}$ increases first at the middle radius ($r_{\rm encl} = 1.0 \Pc$) because of 
gravitational collapse, which is overcome by feedback at $\approx 0.2 \Myr$, reversing the inflow 
and decreasing  $M_{\rm encl}$ also at this radius. 
$M_{\rm encl}$ drops suddenly at $\approx 0.93 \Myr$ when the gas is swept by a partial shell. 
The wealth of various phenomena seen in the "C" models as contrasted to monotonically expanding shell in "O" 
models indicates that resolving the cluster changes substantially the process of gas expulsion. 

The evolutionary sequence for the cluster of mass $M_{\rm embd} = 9\times 10^3 \Msun$ is shown 
in the left two columns of \reff{fse23}. 
Model O9e3 (the leftmost column) drives an ionisation front and wind driven bubble into the cloud, whose 
inner part is swept into a shell (the panel at $0.25 \Myr$).  
The shell has a larger surface density than in the previous case (model O3e3), so it breaks sooner into fragments and filaments. 
Feedback is not able to expel the densest filaments, which start falling inwards and envelop the source, 
quenching its feedback. 
This occurs at $0.42 \Myr$. 
This is also seen on the $M_{\rm encl}(r_{\rm encl} = 0.5 \Pc)$ plot 
(the solid green line on the left central panel of \reff{fgasExp}), 
where the centre is evacuated at $0.1 \Myr$ by the shell, but then swamped by the inflow after $0.4 \Myr$. 
With feedback quenched, the collapse continues with almost all the gas falling to the cluster centre. 

The resolved cluster (model C9e3) again shows more complex behaviour. 
While the stars located near the cloud centre are not able to stop the collapse, 
the stars located further away ionise and ablate its outer parts, which have lower escape speed from the potential of the cluster, unbinding 
approximately $30$\% of the cloud mass by the time of $1 \Myr$ (left central panel of \reff{fgasExp}). 
The majority of the denser inner cloud collapses to a dense structure, which would form stars if star formation 
were allowed in our simulations. 
The majority of the former cloud volume is ionised and dispersed, which is in contrast to the one source model O9e3, 
where the cloud collapses after the single source has been quenched. 

With increasing cluster mass, the 
one source models (O3e4 and O9e4; Figs. \ref{fse23} and \ref{fse4}) show earlier 
quenching of feedback (at $0.1 \Myr$ and $0.02 \Myr$, respectively), 
whereupon the cloud centre is swamped by gravitational collapse 
(the mass within the radii $r_{\rm encl}$ is shown on the centre right and bottom left panel of \reff{fgasExp}). 
As expected, swamping the source and gravitational collapse happens faster for the more massive clouds with 
shorter $t_{\rm ff}$ (c.f. \reff{tListSimsMain}). 
Resolved counterparts to these clouds (models C3e4 and C9e4) also show the decreasing ability of 
feedback to counteract the gravitational collapse; however, they are still able to 
unbind $10$\% to $20$\% of the initial cloud mass and evacuate the majority of the cloud volume. 
The cloud shrinks to a small volume (not seen beyond the stars in Figs. \ref{fse23} and \ref{fse4}), 
which would form stars and disperse under their feedback. 
The rates of gas flow through the sphere of radius $1 \Pc$ are summarised in the lower right panel of \reff{fgasExp}.

Note that the complicated structures seen in "C" models are caused purely by the discretisation of star clusters 
to massive stars because the initial gaseous configuration is spherically symmetric with zero velocity. 
This sets a lower estimate on the complexity of the gaseous component 
in the presence of massive stars; real interstellar clouds are far from being spherically symmetric 
even before the onset of star formation.

\subsubsection{The time-scale of gas expulsion}

%\iffigscl
%\begin{figure}
%\includegraphics[width=\columnwidth]{GEX_tge}
%\caption{\figpan{Left axis:} The gas expulsion time-scale $t_{\rm ge}$ as a function 
%of embedded cluster mass $M_{\rm embd}$ for the "O" models (green circles) and "C" models (red circles). 
%If the majority of gas remained unexpelled by the end of the simulation at $1 \Myr$, this fact is 
%shown by the arrows. 
%The cluster mass for which the escape speed from the cloud centre equals the sound speed in 
%an \HII region is indicated by the yellow line. 
%\figpan{Right axis:} The mean surface density $\overline{\Sigma}_{\rm gas}$ of 
%the cloud inside radius $a_{\rm Pl}$ is shown by the black lines; 
%the black solid line shows the value for $a_{\rm Pl} = 1\Pc$, 
%black dashed line is for a more collapsed cloud with $a_{\rm Pl} = 0.3\Pc$. 
%The surface density threshold of $1.2 \Gcmii$ \citep{Fall2010} for radiation pressure 
%to dominate the other forms of feedback is indicated by the blue line.}
%\label{fgeTimescale}
%\end{figure} \else \fi

\begin{table*}
\begin{tabular}{ccccccccc}
%  & \multicolumn{2}{c}{"O" models} & \multicolumn{2}{c}{"C" models} \\
%$M_{\rm embd}$ & $t_{\rm ge}$ & $\frac{M_{\rm encl}(1 \Pc, 1 \Myr)}{M_{\rm encl}(1 \Pc, 0)}$ & $t_{\rm ge}$ & $\frac{M_{\rm encl}(1 \Pc, 1 \Myr)}{M_{\rm encl}(1 \Pc, 0)}$ 
%& $\overline{\Sigma}_{\rm gas} (1 \Pc)$ & $\overline{\Sigma}_{\rm gas} (0.3 \Pc)$ & $v_{\rm esc}$ & $t_{\rm h}$ \\
%$[10^3 \Msun]$ & [$\Myr$] & & [$\Myr$] & & [$10^3 \Sd$] & [$10^3 \Sd$] & [$\Kms$] & [$\Myr$]  \\
%[$10^3 \Msun$] & [$\Myr$] & & [$\Myr$] & \\
 & & & & & \multicolumn{2}{c}{O models} & \multicolumn{2}{c}{C models} \\
\hline
$M_{\rm embd}$ & $\overline{\Sigma}_{\rm gas} (1 \Pc)$ & $\overline{\Sigma}_{\rm gas} (0.3 \Pc)$ & $v_{\rm esc}$ & $t_{\rm h}$ &
$t_{\rm ge}$ & $\frac{M_{\rm encl}(1 \Pc, 1 \Myr)}{M_{\rm encl}(1 \Pc, 0)}$ & $t_{\rm ge}$ & $\frac{M_{\rm encl}(1 \Pc, 1 \Myr)}{M_{\rm encl}(1 \Pc, 0)}$ \\
$[10^3 \Msun]$ & [$10^3 \Sd$] & [$10^3 \Sd$] & [$\Kms$] & [$\Myr$] & [$\Myr$] & & [$\Myr$] & \\
\hline
% see Exc. book SILCC II from 21st May 2019
%0.9 &  0.26 &  $1 \times 10^{-3}$  &  $> 1$ & 0.65 & 0.1 & 1.1 & 2.8 & 2.4 \\
%3.0 &  0.18 &  $6 \times 10^{-7}$  &  0.9 & 0.05 & 0.32 & 3.5 & 5.1 & 1.3 \\
%9.0 &  x &  2.5  &  x & 1.7 & 0.96 & 11.0 & 8.8 & 0.76 \\
%30.0 &  x &  2.6 & x & 2.0 & 3.2 & 35.0 & 16 & 0.41 \\
%90.0 &  x &  2.7 & x & 2.2 & 9.6 & 110 & 28 & 0.24
%\end{tabular}
0.9  &  0.1 &  1.1 & 2.8 &  2.4 &  0.26 &  $1 \times 10^{-3}$  &  $> 1$ & 0.65 \\
3.0  & 0.32 &  3.5 & 5.1 &  1.3 &  0.18 &  $6 \times 10^{-7}$  &  0.9   & 0.05 \\
9.0  & 0.96 & 11.0 & 8.8 & 0.76 &  x &                    2.5  &  x     &  1.7 \\
30.0 &  3.2 & 35.0 &  16 & 0.41 &  x &                    2.6  &  x     &  2.0 \\
90.0 &  9.6 & 110  &  28 & 0.24 &  x &                    2.7  &  x     &  2.2 
\end{tabular}
\caption{Summary of the embedded cluster models. 
The first to the fifth column indicate the total (combined gaseous and stellar) mass $M_{\rm embd}$, 
the mean initial gas surface density inside radius $1\Pc$ ($\overline{\Sigma}_{\rm gas} (1 \Pc)$) and 
inside radius $0.3 \Pc$ ($\overline{\Sigma}_{\rm gas} (0.3 \Pc)$), escape velocity $v_{\rm esc}$ from the cluster centre,
and the half-mass crossing time $t_{\rm h}$; these values are the same for "O" and "C" models. 
The sixth and seventh column list the gas expulsion time-scale $t_{\rm ge}$ and the relative change 
of the gas mass $M_{\rm encl}(1 \Pc, 1 \Myr)$ which is enclosed within the radius $r_{\rm encl} = 1 \Pc$ at $1 \Myr$. 
The 'x' symbol indicates that ionising radiation and stellar winds are unable to expel the majority of the gas.
The last two columns list the same quantities for the models with resolved clusters ("C" models).
}
\label{tFinal}
\end{table*}

For simplicity, we define the gas expulsion time-scale $t_{\rm ge}$ as 
the time when $M_{\rm encl}(r_{\rm encl} = a_{\rm Pl} = 1 \Pc)$ drops to $1/e$ of its initial value. 
The dependence of $t_{\rm ge}$ on the cluster mass is listed in \reft{tFinal}.
The least massive clusters with one source (model O9e2) have $t_{\rm ge} = 0.26 \Myr$. 
However, representing the cluster with resolved stars (model C9e2) increases $t_{\rm ge}$ substantially. 
From the gradually decreasing $M_{\rm encl}$ and outflows (the upper left panel 
of \reff{fgasExp}), the cluster will eventually evacuate its 
natal gas, but it will take more time than the $1 \Myr$ of the simulations, i.e. $t_{\rm ge} > 1 \Myr$. 
  
With increasing cluster mass to $M_{\rm embd} = 3\times 10^3 \Msun$, $t_{\rm ge}$ decreases to $0.18 \Myr$ for model O3e3 and 
to $0.9 \Myr$ for model C3e3. 
Again, the resolved cluster has substantially longer gas expulsion time-scale than the idealised one source model. 
One source models are also more efficient at clearing the cluster out of the gas as 
it is shown by the ratio $M_{\rm encl}(1 \Pc, 1 \Myr)/M_{\rm encl}(1 \Pc, 0)$ of the initial and final 
gaseous mass within the sphere of radius $1 \Pc$. 
The feedback is less efficient in clusters with $M_{\rm embd} \gtrsim 9\times 10^3 \Msun$, so 
these clusters cannot be evacuated from their natal gas by ionising radiation and stellar winds only. 
%This does not necessarily imply that their natal gas must form more stars to increase feedback to clear the cluster 
%from the gas; instead,  
%radiation pressure (which is absent in our simulations) might be the dominant feedback mechanism in these clusters. 

We discuss the possibility whether the more massive clusters can be evacuated by another feedback mechanism: 
radiation pressure, which is absent in present simulations. 
%To elaborate a bit more on this idea, we compare the 
For this purpose, we compare the 
mean surface density for Plummer models within projected radius $a_{\rm Pl}$
with the gas threshold surface density $\Sigma_{\rm thr} = 1.2 \Gcmii$ ($5800 \Sd$), 
where radiation pressure is likely to be the dominant feedback mechanism \citep{Fall2010}. 
We consider two values of $a_{\rm Pl}$ (c.f. \reft{tFinal}): $1 \Pc$ (at the beginning of the simulation) 
and $0.3 \Pc$ (the smallest radii of observed embedded star clusters). 
Clusters with $M_{\rm embd} \lesssim 3\times 10^3 \Msun$ have initially $\overline{\Sigma}_{\rm gas} (1 \Pc) < \Sigma_{\rm thr}$, 
and their surface gas density decreases (because their volume density decreases as seen in the upper row of \reff{fgasExp}), 
so they are evacuated with ionising radiation and winds only without a significant contribution of radiation pressure. 
Clusters in the mass range $M_{\rm embd} \in (\approx 3\times 10^3 \Msun, \approx 5\times 10^4 \Msun)$ initially 
have $\overline{\Sigma}_{\rm gas} (1 \Pc) < \Sigma_{\rm thr}$, but their surface density increases with time 
as they collapse and when reaching radius $0.3 \Pc$, their surface density exceeds $\Sigma_{\rm thr}$ (\reft{tFinal}), 
so the role of radiation pressure might increase, bouncing the collapse and dispersing the gas. 
Clusters more massive than that ($M_{\rm embd} \gtrsim 5\times 10^4 \Msun$) have feedback dominated by 
radiation pressure from the beginning, and they might expand since this point. 
For these reasons, we cannot estimate $t_{\rm ge}$ for clusters 
of mass $M_{\rm embd} \gtrsim 9\times 10^3 \Msun$ based on present models, 
but the results for $M_{\rm embd} \lesssim 3\times 10^3 \Msun$ seem to be reliable as feedback 
in these clusters is likely dominated by ionisation and wind feedback.

\subsubsection{A note on the star formation efficiency}

Clusters of mass $M_{\rm embd} \lesssim 3\times 10^3 \Msun$ can easily disperse their parent gas cloud if formed with relatively 
low SFE of $1/3$ only with photoionising and stellar wind feedback. 
However, clusters with $M_{\rm embd} > 3\times 10^3 \Msun$ are unable to disperse their natal gas 
with photoionising and stellar wind feedback only if formed with this low SFE. 
This implies two possibilities for clusters with $M_{\rm embd} \gtrsim 3\times 10^3 \Msun$: they 
either continue forming stars (and increase their SFE) until their photoionising feedback is powerful enough to overcome self-gravity; 
or that another form of feedback (e.g. radiation pressure) disperses their parent cloud.

%\begin{itemize}
%\item
%The lowest mass clusters clear their natal gas; the most masive models are swamped. 
%\item
%The threshold for clearing the gas by ionrad + winds is near $M_{\rm cl} = 3\times 10^3 \Msun$ with escape speed $9 \Kms$ - 
%it is close to the condition by Dale. 
%\item
%Although the threshold cluster mass is the same between O and C models, the gas expulsion time-scale is different, 
%and the scenario in C models is more complicated than a simple shell.
%\item
%Although the gas clouds are spherically symmetric, the discrete distribution of massive 
%stars redistributes the cloud so that the symmetry is broken. 
%\item
%C models erode some gas (30\%) also for the more massive clusters C3e4 and C9e4, but this is mainly from their outer parts.
%\item
%O9e4 and O1e5 are swamped by a gaseous inflow and suffocate from the beginning. The short free-fall time 
%at the centre is mainly due to the gravitational attraction of all the sink particles sitting just at one place 
%and constituting substantial mass. 
%This is seen in the small sim in \verb|/home/franta/Desktop/simulations_smallcomputer/flash/sink_tests/GEX/one_source/M1e5|.
%\end{itemize}

\subsection{Discussion}

% Comparison to previous works: Dale2011, Rahner2017, negligence of ram pressure.

%Although our models start with idealised clusters and clouds, the gas expulsion time-scale is close  
%to the analysis of \citet{Pfalzner2019}. 
Based on Gaia observations of young star clusters due to \citet{Kuhn2019}, \citet{Pfalzner2019} 
finds that the gas expulsion time-scale decreases from $\approx 2 \Myr$ to $\approx 1 \Myr$ with increasing cluster mass. 
This is close to our findings in resolved clusters (models C9e2 and C3e3), but not in the 
idealised one source models (O9e2 and O9e3), which have substantially shorter $t_{\rm ge}$ of $\approx 0.2 \Myr$. 
The time-scale $t_{\rm ge}$ for the "O" models is by a factor of $2$ within the sound crossing time $a_{\rm Pl}/c_{\rm II}$, 
where $c_{\rm II}$ is the sound speed in ionised hydrogen (typically $\approx 10 \Kms$), but $t_{\rm ge}$ for the "C" (resolved) models is different 
by a factor of $10$. 
This comparison, which should be taken with caution due to the simplicity of our models, might indicate that 
resolution of star clusters to individual massive stars instead of taking all of them as a single point source 
(as is usually done in star formation simulations) is important for capturing gas 
dynamics in young star clusters.

The gas expulsion time-scale determines (together with the SFE) 
whether the newly born star cluster will survive its emergence from the cloud 
as a gravitationally bound entity or not (e.g. \citealt[][]{Hills1980,Lada1984,Baumgardt2007}). 
The important criterion is the duration of gas expulsion $t_{\rm ge}$ relative to stellar 
half-mass crossing time $t_{\rm h}$. 
If gas expulsion occurs adiabatically ($t_{\rm h} \ll t_{\rm ge}$), the star cluster 
has enough time to react gradually on the change of gravitational potential, 
with substantially less stars unbound 
than if gas expulsion acts impulsively ($t_{\rm h} \gg t_{\rm ge}$), where the 
cluster typically loses more than one half of its stars for SFEs $\lesssim 0.4$. 

We can apply our findings only to clusters with $M_{\rm embd} \lesssim 3\times 10^3 \Msun$, because gas expulsion 
from more massive clusters is likely dominated by radiation pressure, which we neglect in our simulations. 
For the lower mass clusters, gas expulsion due to ionising radiation and winds occurs on 
a time-scale comparable to the half-mass crossing time $t_{\rm h}$ of the cluster (\reft{tFinal}). 
It is likely that real clusters are more concentrated than their host clouds \citep{Parmentier2013}, 
thus having shorter $t_{\rm h}$ than assumed in this work, i.e. $t_{\rm h} \lesssim t_{\rm ge}$. 
If true, this would imply, in contrast to what is often assumed \citep[e.g.][]{Geyer2001,Kroupa2001b}, 
that gas expulsion does not occur impulsively, 
less impacting the cluster dynamics and unbinding a smaller fraction of stars from the cluster. 
This study gives us no information about the nature of gas expulsion 
for clusters of $M_{\rm embd} \gtrsim 3\times 10^3 \Msun$, so we cannot predict whether these clusters have adiabatic or impulsive gas expulsion. 

\citet{Dale2012} find that the approximate condition when ionising radiation overcomes gravity can be formulated 
as the condition when the escape speed $v_{\rm esc}$ from the \HII region is larger than 
its sound speed $c_{\rm II}$. 
In our simulations, ionising and wind feedback becomes inefficient 
for clusters with $M_{\rm embd}$ somewhere in the range of $3\times 10^3 \Msun$ to $9\times 10^3 \Msun$, which
have $v_{\rm esc}$ between $5.1 \Kms$ and $8.8 \Kms$ (\reft{tFinal}). 
This is close to the condition of \citet{Dale2012}.
Since this result holds for both O and C models, 
the critical mass where photoionising and wind feedback stalls is 
independent on the spacial distribution of massive stars within the cloud.

\section{Summary}

\label{sSummary}

To improve the sink particle integrator in hydrodynamic code \flashd, 
we implemented there a 4th order Hermite predictor-corrector scheme with individual 
quantised particle block time steps.
The integrator splits the fast-changing and slow-changing force component by the Ahmad-Cohen method; 
however in contrast to the standard Ahmad-Cohen method, we base the split on the kind of the mass which 
causes the force; sink particles always belong to the fast-changing force component while 
gas always belongs to the slowly changing force component. 

We present an extensive suite of tests to compare this integration scheme with the 
current sink particle integrator in \flash (leap-frog scheme) as well as with 
the state-of-the-art N-body integrator \nbdvid. 
We perform both tests with negligible density of the background gas to test pure stellar dynamics as well as tests 
containing massive gaseous bodies to test the interaction between sink particles and gas. 

In order to model the dynamics of massive stars, which plays role on the time-scale of the 
lifetime of star forming regions, the code admits two types of sink particles: sink particles of the first type  
represent massive stars with a compact softening radius $r_{\rm soft,ss,1}$, which
allows accurate modelling of close stellar encounters; 
sink particles of the second type 
represent lower mass stars with a generous softening radius $r_{\rm soft,ss,2}$. 
We demonstrate that increasing the softening radius for lower mass stars and reducing their number 
has a negligible influence on dynamics of young star clusters (\reff{fHalfMass}), but enables to 
accurately calculate close encounters between massive stars, which mass segregate to the cluster centre.
For practical purposes, it is possible to directly calculate small systems of strongly interacting massive stars, 
which self-consistently form hard binaries and fast ejectors with properties very close to that 
obtained from \nbdvi (\reff{fScatter}).

In the pure stellar dynamical simulations, the Hermite scheme is faster than the leap-frog \flash integrator 
by a factor of $10^3$ (in CPU time) for the same or better accuracy as measured by the error 
in mechanical energy conservation. 
An increase in both accuracy and 
performance in comparison to the leap-frog is also observed in simulations containing 
both sink particles and massive gaseous bodies (\reff{fBESphere_2p}).
The integrator is able to model a compact $N = 10^4$ star cluster for 
ten half-mass radius crossing times in $\approx 10^5$s of CPU time, making such an integration feasible 
in modern hydrodynamic simulations. 
Current implementation of the Hermite scheme preserves all the criteria of \citet{Federrath2010} 
for sink particle creation and accretion. 
We plan to make the integrating module, which we call \textsc{Hermitesink}, 
to be a publicly available part of code \flashd.

We employ the Hermite integration scheme to study the gas expulsion process from embedded star clusters. 
Initially, the embedded clusters have total mass 
in the range of $M_{\rm embd} = 9\times 10^2 \Msun$ to $M_{\rm embd} = 9\times 10^4 \Msun$, 
where mass $M_{\rm embd}/3$ is in the form of stars, and $2M_{\rm embd}/3$ in the form of gas. 
This means that we assume initial SFE of $1/3$ with no star formation occuring in the course of the simulation. 
Massive stars are modelled individually, and they are distributed within 
a Plummer sphere of Plummer scale-length $1 \Pc$, 
and subjected to their mutual gravitational encounters, the attraction of lower mass stars, 
as well as the live gravitational field of the gas. 
We consider feedback from massive stars in the form of ionising radiation and stellar winds. 
The simulations terminate at $1 \Myr$, well before the occurrence of the first supernova.

The models show that embedded star clusters with $M_{\rm embd} \lesssim 3\times 10^3 \Msun$ can expel 
all of their natal gas only by ionising radiation and stellar winds even if the clusters formed with 
relatively low $\sfe = 1/3$. 
On the other hand, clusters with $M_{\rm embd} \gtrsim 3\times 10^3 \Msun$ cannot expel 
their natal gas only by the combined effect of ionising radiation and stellar winds for $\sfe = 1/3$. 
This implies that either feedback in these clusters is dominated by another physical mechanism (e.g. radiation pressure), 
or that these clusters continue forming stars (and thus increasing their SFE) until the stellar feedback manages 
to disperse the decreasing supply of their residual gas. 

We test limitations of approximating the star cluster by a single source instead of resolving the 
cluster to individual stars, which is often adopted in hydrodynamic simulations.
For lower mass clusters ($M_{\rm embd} \lesssim 3\times 10^3 \Msun$), 
this approximation leads to a short gas expulsion time-scale ($t_{\rm ge} \approx 0.2 \Myr$), with gas being swept in a single shell.  
Resolved star clusters have significantly longer gas expulsion time-scale ($t_{\rm ge} \gtrsim 0.9 \Myr$), 
where $t_{\rm ge}$ slightly decreases with cluster mass, 
%bringing the models closer to observations \citep{Kuhn2019,Pfalzner2019}. 
bringing the models closer to observations. 
The resolved clusters also show more complex gas morphology with a network of intersecting shells and filaments; the gas is not swept in a single shell.
The complicated morphology develops from clouds which are initially spherically symmetric, with only massive stars having a discrete distribution. 
This indicates that placement of sources of feedback can be of importance comparable to the initial conditions for gas.
We note that the simulations provide us with the upper estimate of the gas expulsion time-scale 
as they neglect other feedback mechanisms (e.g. radiation pressure).

For more massive clusters ($M_{\rm embd} \gtrsim 3\times 10^3 \Msun$), 
the single source approximation results in early quenching of the feedback without any 
noticeable impact on the 
cloud, while resolved clusters are still able to unbind $10$\% to $30$\% of the cloud mass and ionise 
the majority of the cloud volume by $1 \Myr$ mainly by photoionisation of the outer layers of the cloud. 
This indicates that the influence of stellar feedback on the surrounding gas sensitively depends on 
positioning stars within the cluster, and that resolving the cluster to individual massive stars 
is necessary to model the gas expulsion process more realistically.

%--------------------------------------------------------------------
\section*{Data availability statement}

Data available on request.

%--------------------------------------------------------------------
\section*{Acknowledgments}

FD would like to thank to Sverre Aarseth for many discussions about various N-body algorithms. 
%FD and SW thank the DFG for funding through the Collaborative Research Center (SFB956) on 
%the "Conditions and impact of star formation", sub-project C5.
FD and SW acknowledge the support by the Collaborative Research Centre 956, 
sub-project C5, funded by the Deutsche Forschungsgemeinschaft (DFG) – project ID 184018867.
SW is grateful for support via the ERC Starting Grant no. 679852 (RADFEEDBACK). 
The simulations were calculated on SuperMUC at the Leibniz-Rechenzentrum Garching 
under project "pr94du" (PI Daniel Seifried).

\bibliographystyle{mn2e}
\bibliography{hermite}
  
\bsp

%%%%%%%%%%%%%%%%%%%%%%%%%%%%%%%
\appendix
\section{The softening kernel}

\label{sAppendixA}

% \franta{Check the equations once again before submitting.}

In our implementation to the \textsc{HermiteSink} module, we adopt the 
softening kernel from \citet[][their eq. 21]{Monaghan1985}. 
This softening kernel is generated by mass distribution in the form of
\begin{equation}
\rho(r) = \frac{8 m_{\rm kern}} {\pi r_{\rm kern}^3} \times
\begin{cases}
1 - 6q^2 + 6q^3 ,&  \,\, q \in (0, 1/2) \\
2 (1 - q)^3, &  \,\, q \in (1/2, 1) \\
0, &  \,\, q > 1 \\
\end{cases}
\label{eKernDens}
\end{equation}
where $m_{\rm kern}$ is the kernel mass, $m_{\rm kern} = 4 \pi \int_{0}^{r_{\rm kern}} \rho(r) r^2 \dd r$, and 
$q = r/r_{\rm kern}$ ($r_{\rm kern}$ is equal to the softening radius $r_{\rm soft}$ used above) 
is a proxy for the distance from the centre of the kernel. 

This mass distribution generates a gravitational potential of
\begin{equation}
\Phi(r) = \frac{G m_{\rm kern}}{r_{\rm kern}} \times
\begin{cases}
16 \left\{ \frac{1}{3} q^2 - \frac{3}{5} q^4 + \frac{2}{5} q^5  \right\} - \frac{42}{15}, & \,\, q \in (0, 1/2) \\
4 \left\{ \frac{8}{3} q^2 - 4q^3 + \frac{12}{5} q^4 - \frac{8}{15} q^5 + \frac{1}{60 q}  \right\} - \frac{16}{5}, & \,\, q \in (1/2, 1) \\
- \frac{1}{q}. & \,\, q > 1 \\
\end{cases}
\label{eKernPot}
\end{equation}
An important property of this kernel is 
that its potential is identical to the Kepler potential, which has $\Phi(r) \propto -1/r$, at $r > r_{\rm kern}$, 
so the dynamics of stars is calculated exactly unless they approach each other 
closer than $r_{\rm kern}$. 
This is not the case, for example, of the Plummer softening potential, which has $\Phi(r) \propto (1+(r/r_{\rm kern})^2)^{-1/2}$, 
and therefore it differs from the Kepler potential at $r > r_{\rm kern}$, 
and transitions to the Kepler potential only at $r \gg r_{\rm kern}$.

The gravitational acceleration $\mathbf{a}$ and its derivative $\dot{\mathbf{a}}$ can be expressed as 
\begin{subequations}
\begin{align}
\mathbf{a}(\mathbf{r}) &= a(r) \frac{\mathbf{r}}{r}, \\
\dot{\mathbf{a}}(\mathbf{r}) &= \left\{ \pder{a(r)}{r} - \frac{a(r)}{r} \right\} 
\frac{(\mathbf{r} \cdot \mathbf{v}) \mathbf{r}}{r^2} + \frac{a(r) \mathbf{v}}{r},
\end{align}
\label{eKernAccaAccDer}
\end{subequations}
where 
\begin{equation}
a(r) = - \frac{G m_{\rm kern}}{r_{\rm kern}^2} \times
\begin{cases}
4 \left\{ \frac{8}{3} q - \frac{48}{5} q^3 + 8 q^4  \right\}, & \,\, q \in (0, 1/2) \\
4 \left\{ \frac{16}{3} q - 12 q^2 + \frac{48}{5} q^3 - \frac{8}{3} q^4 - \frac{1}{60 q^2}  \right\}, & \,\, q \in (1/2, 1) \\
\frac{1}{q^2}, & \,\, q > 1 \\
\end{cases}
\label{eKernAcc}
\end{equation}
and

\begin{equation}
\pder{a(r)}{r} = - \frac{G m_{\rm kern}}{r_{\rm kern}^3} \times
\begin{cases}
32 \left\{ \frac{1}{3} - \frac{18}{5} q^2 + 4 q^3  \right\}, & \,\, q \in (0, 1/2) \\
32 \left\{ \frac{2}{3} - 3 q + \frac{18}{5} q^2 - \frac{4}{3} q^3 + \frac{1}{240 q^3} \right\}, & \,\, q \in (1/2, 1) \\
- \frac{1}{q^3}. & \,\, q > 1 \\
\end{cases}
\label{eKernAccDer}
\end{equation}

\label{lastpage}

\end{document}